# Hysteresis and Duration Dependence of Financial Crises in the US: Evidence from 1871-2016

**Rui Menezes**[a,c]

**Sonia Bentes**[b]

[a] ISCTE-IUL and BRU-IUL

[b] ISCAL and BRU-IUL

[c] corresponding author: rui.menezes@iscte.pt

**31 August 2016**

**Abstract**

This study analyses the duration dependence of events that trigger volatility persistence in stock markets. Such events, in our context, are monthly spells of contiguous price decline or negative returns for the S&P500 stock market index over the last 145 years. Factors known to affect the duration of these spells are the magnitude or intensity of the price decline, long-term interest rates and economic recessions, among others. The result of interest is the conditional probability of ending a spell of consecutive months over which stock market returns remain negative. In this study, we rely on continuous time survival models in order to investigate this question. Several specifications were attempted, some of which under the proportional hazards assumption and others under the accelerated failure time assumption. The best fit of the various models endeavored was obtained for the log-normal distribution. This distribution yields a non-monotonic hazard function that increases up to a maximum and then decreases. The peak is achieved 2-3 months after the spell′s onset with a hazard of around 0.9 or higher; this hazard then decays asymptotically to zero. Spell′s duration increase during recessions, when interest rate rises and when price declines are more intense. The main conclusion is that short spells of negative returns appear to be mainly frictional while long spells become structural and trigger hysteresis effects after an initial period of adjustment. Although in line with our expectations, these results may be of some importance for policy-makers.

JEL classification: C41 C58 G10 G15

*Keywords*: Hysteresis, duration dependence, negative returns, survival models, hazard rate

## Table of Contents

# 1 Introduction

This study deals with the duration of spells of stock market price decline in the United States using monthly data from the S&P500 over the past 145 years. The original data belong to a monthly time series available online in Robert Shiller's home page. On the basis of real prices, we computed stock market price returns and retained the information on all monthly episodes of negative returns. Spells were obtained for sequences of adjacent episodes of price decline and the end of a spell occurs when the event of interest takes place. The event of interest is, in our study, the beginning of any episode of no price decline.

The main objective of the study is to identify why some spells of price decline are short while others persist for relatively long periods. In this context, it is important to verify whether there is any relationship between the length of an ongoing spell and its probability of ending in the next time period and how does this pattern evolve over time. In fact, this is important insofar the persistence of a stock market crisis has severe effects on the economy and social welfare in general, and persistence may trigger chronic crises with effects similar to hysteresis.

Although persistence and hysteresis are very similar concepts, the latter is usually associated to situations where the process under study acquires 'memory' and does not respond anymore to external stimulus designed to remove it. That is the properties of the process become endogenous and resilient to unsuitable external factors.

One possible way to study this phenomenon is to rely on survival analysis and duration models. Some studies adopt a discrete time structure based on logit-type regression, although a continuous time framework can also be used when the time window is sufficiently large. Clearly this is our case and we then adopt the framework of continuous time survival models. Several specifications were attempted and a log-normal was selected on the basis of choice criteria. Other precautions related to the correct estimation of the models were also taken into account. The results are robust to the hypothesis of hysteresis, in the sense of negative duration dependence, but only after a short initial period of increasing hazard.

The remain of the manuscript is organized as follows. In section 2 we provide the background of the study, explaining the role of stock market volatility and the concepts of persistence and hysteresis. Section 3 presents the methodological issues of the study, defining the concepts of survival and hazard function, the use of parametric distributions, extreme value parameterization and several types of regression models. We also discuss the different uses and interpretation of proportional hazards and accelerated failure time models. The section ends with a discussion of the likelihood function and some caveats concerning estimation issues such as robust estimation and frailty. In section 4 we describe the dataset used in this study. First, we characterize the economic recessions and financial market breakdowns that occurred in the US over the last 145 years. Next we describe the behavior of the S&P500 stock market returns over the entire span. A subsection dedicated only to asset price declines is presented next, within which there is another subsection explaining the spell duration of asset price declines. A description of the choice of relevant covariates is given in the next subsection and the section ends with a presentation of the survival time and failure rates of price declines. Section 5 presents and discusses the main results obtained from estimating the models described in section 3. First we discuss the choice between proportional hazards and accelerated failure time models. Next, among the several parametric models, the best fit is discussed in the subsection destined to estimation. A discussion of the estimated parameters (covariates and ancillary) is offered in 5.2.1, followed by the shape and interpretation of the hazard functions for the full model. The final subsection is reserved to discuss the shape and interpretation of the hazard functions for several stratifications given by the quantiles of the continuous time covariates as well as the link of these patterns with the concept of hysteresis. The results are also separated according to a factor dummy denoting the occurrence of US recessions. Finally, the main conclusions are summarized in section 6.

## 2 Background

### 2.1 Volatility in Stock Markets

In finance, the term asset price volatility is often used to describe the rate at which the price of an asset moves up and down. In general, it is associated with the notions of variability, zigzagging, unpredictability, uncertainty or risk, although the last one is perhaps the most vulgarized synonymous. High volatility is commonly meant as a sign of market disruption, instability or turbulence, i.e. a state of confusion and market malfunctioning without any order. When associated with unpredictability it describes the degree of dispersion around an expected value. Uncertainty and risk have a probabilistic interpretation whereby price variation is randomly determined by a probability distribution but may not be foreseen precisely. Anyway, uncertainty and risk are not exactly the same (Knight, 1921). While the term risk means that one does not know with certainty the result of a decision or process but its probability distribution is known *a priori*, the probability distribution in a situation of uncertainty is unknown. In this way, the risk can be seen as a quantification of uncertainty. Therefore, one should define as accurately as possible the scope of the term volatility used in the context of the study being carried out (Dionísio *et al.*, 2007).

Granger (2005) offers a different explanation for the distinction between risk and uncertainty, where risk is associated with negative shocks and uncertainty is related with positive shocks. For instance, for a given portfolio in which an asset faces a large negative shock, the probability that the asset is traded at a lower price than the buying one, thus incurring into a loss, increases and this is called risk, or the risk of loss. However, if the shock is positive, this is understood as uncertainty, or the uncertainty of gaining. Thus, although both shocks yield increases in the variance, only the negative shock is seen as undesirable and this is the rational for distinguishing both concepts. Risk and uncertainty are therefore two different faces of the same coin.

Investors can be classified into different types according to their attitude facing uncertainty and risk as aggressive (speculator), moderate (risk-neutral) or conservative (risk-averse). Since rational investors maximize their expected profits or returns subject to uncertainty/risk constraints, the tradeoff between uncertainty/risk and expected return is central to many modern theories in finance. Notice that the higher is the uncertainty/risk, the higher is not only the expected maximum gain but also the expected maximum loss. It is therefore a problem of statistics of extremes. But while expected gain is not controversial among the different types of investors, the expected loss may trigger different decisions between, for example, speculators and risk averse investors. Hedging strategies have been devised in order to minimize the negative impact of expected losses among risk averse investors. However, these strategies may not be sufficiently attractive for conservative investors, who will still react negatively to substantial changes in the volatility of expected returns (Fama, 1970; Lansing and LeRoy, 2014).

Volatility may arise from systematic deviations from the steady state market path, given the effect of market imperfect information, or from irregular deviations that occur occasionally. It is well known that, for example, an increase in stock market volatility is associated with an increased probability of large returns of either sign, but prices themselves tend to move down (Black, 1976; Christie, 1982; Schwert, 1989). This is also the case with our dataset as we shall see later in this study. Large returns of either sign occurred, for instance, during the Great Depression of the 1930s and the subprime crisis of 2008.

Daly (2008) points out six reasons to highlight the importance of studying volatility in financial markets:

1. Sharp price changes over short time spans (e.g. intraday) are not explained by qualitative or quantitative economic fundamentals leading to reduced flows of capital and lower liquidity in equity markets;
2. The probability of bankruptcy increases with volatility given a relatively stable asset structure of a firm because the equity value reflects the market price of the shares and default becomes more likely;
3. The bid-ask spread increases with volatility because risk premium rises and liquidity falls;
4. Hedging strategies are affected by volatility because insurance prices increase;
5. If consumers are mostly risk averse, an increased risk associated with a given economic activity tends to reduce the level of participation in that activity with adverse consequences for investment;
6. During periods of high volatility, the regulatory agencies may force firms (e.g. banks) to reallocate capital to cash-equivalent investments thus reducing the allocation efficiency and the value of firms.

Nelson (1990), on the other hand, identifies four stylized facts usually associated with changes in volatility over the long-term:

1. Large changes in volatility tend to be followed by large changes of either sign and vice-versa, i.e. positive serial correlation or volatility clustering (Mandelbrot, 1963a);

2. When prices fall they become more leveraged and the volatility of price changes rises, i.e. there are partial leverage effects (Schwert, 1989);

3. Volatility tends to be high during recessions and financial crises (Black, 1976; Christie, 1982; French *et al.*, 1987; Schwert, 1990);

4. High nominal interest rates are associated with high volatility (Christie, 1982; Fama and Schwert, 1977).

In addition, trading volume, contrarian trade, trading/non-trading days, and futures and options are all factors associated with changes in volatility over the short-term (Fama, 1965; French and Roll, 1986).

Many measures of stock market volatility have been proposed in the literature but perhaps the most commonly used is the standard deviation of returns. Returns, on the other hand, are not directly observed but can be obtained by $r_t = \ln P_t - \ln P_{t-1}$, where $P$ denotes the price and $t$ is the moment of measurement (day, week, month, etc.). In our study we are only interested in negative real price changes, i.e. in the words of Granger (2005), the changes associated with risk, and so volatility is measured as the percentage change in prices from $t–1$ to $t$ when price change is negative, denoting the intensity of the price decline.

## 2.2   Persistence and Hysteresis

The term persistence is used when someone or something persists, i.e. turns to be resilient in their path. In our context we can say that volatility persists when it tends to resist over time at a given level or within a given interval. This may be a problem if we are thinking of high levels of volatility, that is, an undesirable situation for risk-averse investors.[1] High levels of volatility are sometimes referred to as excess volatility but the threshold above which there is excess of volatility is not necessarily defined *a priori* (see e.g. Bartolini and Giorgianni, 1999; Giglio and Kelly, 2015; LeRoy, 2005 for a debate on excess volatility). The phenomenon of persistence is also known as volatility clustering (Bentes *et al.*, 2008) and is usually detected by long-lasting positive serial correlation, whereby large fluctuations in volatility tend to be followed by further large fluctuations and vice-versa, as noted above.

---

[1] Although it may be a 'heaven' for speculators.

A persistent process is a process with long-term memory and thus both concepts are in some way linked. Hurst exponents, for example, are commonly used to detect the extent of memory in a time series. The $H$ exponent ranges between 0 and 1 and measures three types of trends: persistence or long-memory ($H > 0.5$), randomness or no autocorrelation ($H = 0.5$) and anti-persistence or mean reversion ($H < 0.5$). Note that $H > 0.5$ is also indicative of long-term positive autocorrelation and $H < 0.5$ indicates a time series with long-term switching between high and low values in adjacent pairs (Makarava *et al.*, 2014; Mandelbrot, 1963b; Mitra, 2012; Rasheed and Quian, 2004). Alternative measures for detecting the extent of memory in a time series have been proposed in the literature. One such measure is based on the fractional exponent of a FIGARCH-type model as proposed by Baillie *et al.* (1996). The $d$ parameter plays here a similar role as the $H$ exponent. So, the longer is the memory of the process the higher is its persistence and the more time a shock takes to vanish.

Another important notion in the context of our study is hysteresis. Hysteresis denotes the persistence in long-run trends in volatility. In physics, for example, when a material becomes magnetized by an external magnetic field, it will stay magnetized indefinitely until an opposite source is applied to that material.[2] In economics it denotes a setting where some usually undesirable phenomenon persists, such as for instance long-term unemployment.[3] In financial markets hysteresis is a situation that endures despite evidence that it should not, eventually because of the delayed effects of some past event that continue to affect the future. Any disturbance in a financial system will lead to a trickle-down effect, and the problem may persist for long due to market rigidity. This rolling down impact is known as the hysteresis effect and we may think of it as a collective memory of the system, of which the consequences of the herding behavior can be an example (see, e.g. Choi, 2016; Economou *et al.*, 2016; Litimi *et al.*, 2016; Rülke *et al.*, 2016)

---

[2] Hysteresis is a Greek term coined by Sir J. Ewing to explain the delay between the action and reaction of a physical device.

[3] In economics a commonly cited example of hysteresis refers to negative duration dependence. The longer a person is unemployed the lower are her chances to get new employment because: *i*) as time passes unemployed workers lose part of their skills and become less attractive to potential employers, *ii*) market stigmatization, i.e. a sign of disgrace associated with unemployment or the unemployed, *iii*) the unemployed may adjust to a lower standard of living and are not so determined in attaining their previous status, *iv*) some unemployed people may become discouraged about the prospects of getting new employment and leave the work force. This results in higher structural unemployment (Blanchard and Diamond, 1990; Heckman and Borjas, 1980).

In reality, an example of hysteresis in financial markets is the persistence of volatility over time where, e.g. a spell of high volatility tends to be followed by another spell of high volatility and vice-versa, even if the causes of the crisis have already vanished. As mentioned above, this is known in the financial literature as volatility clustering or persistence, indicating the delayed effect of certain measures to promote recovery following a crisis. Hysteresis is therefore linked to the duration dependence of spells of high volatility with returns of either sign but if we are concerned with risk we may just concentrate on the duration of negative returns (price decline).

When a financial crisis begins one does not know exactly how long it will last but, from historical information, there is a probability distribution that it will last a certain time interval, given a set of explanatory variables. This probability distribution may be of different shapes and there are cases where the probability is constant over time, increases, decreases, or both. When the probability increases over time we say that there is positive duration dependence and negative duration dependence occurs otherwise (van den Berg, 1994; Zorn, 2000). In the latter case, the crisis tends to persist as time progresses with the consequences resulting from its nature. Obviously, negative duration dependence associated with a financial crisis has always a negative connotation so, understanding why and when negative duration dependence occurs is of crucial importance to investors and other decision makers (Aaberge, 2002). Further details about hysteresis can be found in Preisach (1935), Szabó (2006) and Visintin (1995).

## 3 Methodology

### 3.1 Survival Distribution, Hazard Function, Integrated Hazard

**Definition 1.** Let $T$ be a non-negative random variable denoting the time until some event of interest, where $F(t)$ is the distribution function and $f(t)$ is the underlying density function, which is not necessarily a set of probabilities (Jenkins, 2005). Because $T$ is non-negative and denotes the elapsed time until an event, the survivor function is given by

$$S(t) = P(T > t) = \int_t^\infty f(t)dt \qquad (1)$$

for $t > 0$. $S(t)$ denotes the probability of "surviving" beyond time $t$, where surviving means that the event of interest has not yet occurred by that time. The survivor function is equal by definition to $1 - F(t)$ and is a monotone non-increasing function of $t$.

Examples of this type of variables include the time from diagnosis of a disease until death (biomedicine), the time interval between stock transactions in the stock exchange (finance), the time until a machine or system fails (engineering), and so on. In the first example, the event of interest is death caused by the disease, in the second the event of interest is the occurrence of a stock transaction and in the third the event of interest is the machine or system failure. $T$ is also referred to as a duration random variable.

**Remark 1.** For an arbitrary survival time $T$, the nonparametric estimators of the distribution and survivor functions, $\hat{F}(t)$ and $\hat{S}(t)$ are right continuous.

**Remark 2.** For a sufficiently high value of $t$, say $t \to \infty$, the survivor function indicates the tail probability of $T$.

**Definition 2.** The instantaneous rate at which spells are completed and the event occurs at time $t$, given that they lasted until $t$, is called the hazard function and is given by

$$\lambda(t) = \lim_{\Delta t \to 0^+} \{P(t \le T < t + \Delta t \mid T \ge t)/\Delta t\} \qquad (2)$$

Because $\lambda(t)$ is a conditional density, where the numerator denotes the unconditional density function and the denominator is the survivor function, the hazard function can also be written as

$$\lambda(t) = \frac{f(t)}{S(t)} = -\frac{d \ln S(t)}{dt} \tag{3}$$

where $S(t) = \lim_{x \to t^-} S(x)$. Hence, the hazard function is the negative logarithmic derivative of the survivor function.

**Remark 3**. Note that $0 \leq \lambda(t) < \infty$, where the bounds mean no risk and certainty occurrence of the event of interest at time $t$; $\lambda(t)$ describes the failure experience of the subject.

**Definition 3.** The integrated hazard function is given by

$$\Lambda(t) = \int_0^t \lambda(u) du = -\ln S(t) \tag{4}$$

for $t > 0$. $\Lambda(t)$ denotes the cumulative rate at which spells are completed at time $t$, given that the event in question has not yet occurred prior to time $t$, i.e. the total risk of failure accumulated up to time $t$. It follows immediately that $S(t) = \exp\{-\Lambda(t)\}$.

In duration analysis the hazard function is, in general, a better descriptor of the phenomenon than the density function, since it gives information about the chance of ending a spell that has been resilient until then. This is so with or without censored time observations. On the other hand, since the hazard rate is a function of time, one may test alternative distribution functions to fit the data, where the hazard rate may be constant or not constant over time, depending on the distribution shape and other parameters.

## 3.2 Parametric Survival Distributions

**Definition 4.** Let $\Gamma(k)$ denote the gamma function where the real part of the complex number $k$ is positive, given by

$$\Gamma(k) = \int_0^\infty u^{k-1} e^{-u} du \tag{5}$$

The above improper integral is the Euler integral of the second kind, which converges absolutely, and is the continuous counterpart of the factorial shifted function $(k-1)!$, $k \in \mathbb{N}$. For $k \in \mathbb{N}$, it should be noted that using integration by parts and solving recursively leads to

$$\int u^{k-1}e^{-u}du = -e^{-u}(k-1)!\sum_{i=1}^{k}\frac{u^{k-i}}{(k-i)!} \tag{6}$$

For $k = 1$ the above expression reduces to the exponential of the form $\{-\exp(-u)\}$. The incomplete gamma function, on the other hand, is given by

$$\gamma(t,k) = \int_{0}^{t}u^{k-1}e^{-u}\,du \tag{7}$$

where $\Gamma(k) = \lim_{t\to\infty}\gamma(t,k)$.

**Definition 5.** The generalized gamma distribution (GG) with parameters $\lambda$, $p$ and $k$ is represented by the following density function (Stacy, 1962)

$$f(t;\lambda,p,k) = \frac{d\gamma\{(\lambda t)^{p},k\}/dt}{\Gamma(k)} \tag{8}$$

where $d\gamma\{(\lambda t)^{p},k\}/dt = \lambda p(\lambda t)^{pk-1}\exp\{-(\lambda t)^{p}\}$ is the first derivative of the incomplete gamma function and $k$ is the shape parameter. The parameters $\lambda$ and $p$ are, respectively, the location and scale parameters. Note that $t,\lambda,p,k > 0$.

**Definition 6.** The survivor function of the generalized gamma distribution is given by

$$S(t;\lambda,p,k) = 1 - \frac{\gamma\{(\lambda t)^{p},k\}}{\Gamma(k)} = \int_{(\lambda t)^{p}}^{\infty}u^{k-1}e^{-u}du/\Gamma(k) \tag{9}$$

where $\Gamma(k) = (k-1)!$ if $k$ is a positive integer and a function of $\sqrt{\pi}$ if $k$ is a non-integer. The hazard and the cumulative hazard functions can be easily computed from the above expressions. Likewise, if $T$ has a generalized gamma distribution then the $r^{\text{th}}$ moment of $T$ is given by $\text{E}(T^{r}) = (1/\lambda^{r})\,\Gamma(k+r/p)/\Gamma(k)$. Therefore, the mean, the variance and other relevant statistics of $T$ can be obtained replacing $r$ by the corresponding order. The expected value and the variance of $T$ are given by

$$\text{E}(T) = \frac{1}{\lambda}\left[\frac{\Gamma(k+1/p)}{\Gamma(k)}\right] \tag{10}$$

$$\text{var}(T) = \frac{1}{\lambda^{2}}\left[\frac{\Gamma(k+2/p)}{\Gamma(k)} - \frac{\Gamma(k+1/p)}{\Gamma(k)}\right]^{2} \tag{11}$$

**Remark 4.** The generalized gamma distribution (GG) is a general (or limit) case of many interesting survival distributions: the gamma when $p = 1$, the Weibull when $k = 1$, the exponential when $p = 1$ and $k = 1$, the log-normal when $k \to \infty$, and so on. It is also related to the Gompertz-Makeham (or log-Weibull) distribution and can be seen as well as a special case of the generalized $F$ distribution.

Substituting the value of the parameters in the above expressions we obtain the mean and variance of each special case distribution, i.e. the exponential, the Weibull, the gamma, and so on. For example, the mean of the gamma distribution is $k/\lambda$ and the variance is $k^2(k/\lambda)^2$, if $k \in \mathbb{N}$.

**Remark 5.** The exponential distribution is memoryless and has constant hazard rate $\lambda(t) = \lambda$. The hazard rate of the Weibull distribution rises monotonically if $p > 1$, is constant if $p = 1$ (exponential) and declines monotonically if $p < 1$. The gamma hazard rate increases monotonically if $k > 1$, from 0 to a maximum of $\lambda$, decreases monotonically, from $\infty$ to an asymptotic value of $\lambda$ if $k < 1$, and is constant if $k = 1$ (exponential). If $k$ is an integer greater than 1 we have the Erlang distribution as a special case of the gamma. In the case of the log-normal distribution, the hazard function increases from 0 to a maximum and then decreases monotonically, approaching 0 as $t \to \infty$.

When the hazard rate $\lambda(t)$ is not constant over time we say that there is duration dependence. This can be monotonic (e.g., Weibull, Gompertz, gamma) or non-monotonic (e.g., log-normal). Positive duration dependence occurs when the hazard rate increases over the duration of a spell and vice-versa for negative duration dependence. Recall that the hazard rate is the conditional probability of ending a spell given that it lasted until then. When the hazard rate is non-monotonic, negative duration dependence follows positive one after the hazard rate reaches a maximum at some point in time. Negative duration dependence is a sign of structural rather than frictional problems in the event of interest and is often referred to as hysteresis or duration persistence (Kiefer, 1988; Preisach, 1935; Szabó, 2006; Visintin, 1995). As discussed in section 2.2 negative duration dependence can lead to hysteresis effects.

### 3.3 Extreme-Value Parameterization

**Definition 7.** Let $X$ be a continuous random variable that has a generalized extreme-value distribution, defined in $\mathbb{R}$, with parameters $\mu$, $\sigma$ and $\xi$, where the cumulative distribution function is represented by

$$F(x; \mu, \sigma, \xi) = \left\{ \exp\left\{ -\left[ 1 + \xi\left(\frac{x-\mu}{\sigma}\right) \right]^{-1/\xi} \right\} \right\} / \Gamma(\xi) \tag{12}$$

for $1 + \xi\,(x - \mu)/\sigma > 0$. In the GEV distribution $\mu, \xi \in \mathbb{R}$ and $\sigma \in \mathbb{R}^+$ (Markose & Alentorn, 2005).

**Definition 8.** The density function of the generalized extreme-value distribution is given by

$$f(x; \mu, \sigma, \xi) = \frac{(1/\sigma)}{\Gamma(\xi)} \left[ 1 + \xi\left(\frac{x-\mu}{\sigma}\right) \right]^{(-1/\xi)-1} \exp\left\{ -\left[ 1 + \xi\left(\frac{x-\mu}{\sigma}\right) \right]^{-1/\xi} \right\} \tag{13}$$

again for $1 + \xi\,(x - \mu)/\sigma > 0$. When $\xi \to 0$, we have

$$f(x; \mu, \sigma, 0) = (1/\sigma) \exp\left( -\frac{x-\mu}{\sigma} \right) \exp\left\{ -\exp\left( -\frac{x-\mu}{\sigma} \right) \right\} \tag{14}$$

This is known as the Gumbel or type I extreme value distribution (Martins & Stedinger, 2000).

**Remark 6.** In the generalized extreme-value distribution, the parameter $\xi$ governs the tail behavior of the distribution. For example, if $\xi > 0$, we have the Fréchet or type II extreme value distribution and if $\xi < 0$, the sub-family is known as the reversed Weibull or type III extreme value distribution. The GEV tends to a generalized Gaussian distribution when $\xi \to -1/2$.

The expected value and the variance of $X$ are given by

$$\mathrm{E}(X) = \mu + \frac{\sigma}{\xi}\left[ \Gamma(1-\xi) - 1 \right] \tag{15}$$

$$\mathrm{var}(X) = \left(\frac{\sigma}{\xi}\right)^2 \left\{ \Gamma(1-2\xi) - [\Gamma(1-\xi)]^2 \right\} \tag{16}$$

where $\Gamma(1 - r\xi)$ denotes the gamma function. The above expressions are valid if $\xi \neq 0$ and $\xi < 1$ (mean) or $\xi < 1/2$ (variance). If $\xi = 0$, $\mathrm{E}(X) = \mu + \sigma\gamma$ and $\mathrm{var}(X) = \left(\sigma\pi/\sqrt{6}\right)^2$; $\gamma$ is the Euler's constant. If $\xi \geq 1$ ($\xi \geq 1/2$) the mean (variance) is infinite.

The GEV is often known as the Fisher-Tippett distribution. The random variable $X$ can be seen as a Lévy process if the increments are independent and stationary and $\lim_{\Delta \to 0} P(|X_{t+\Delta} - X| > \epsilon) = 0; \epsilon > 0, t \geq 0$.

**Definition 9.** Let us now consider the random variable $W$ defined in $\mathbb{R}$ that has a generalized extreme-value distribution and is linearly related to the random variable $T$, defined in $\mathbb{R}^+$, that has a generalized gamma distribution. This linear relation is represented by

$$Z = \ln T = \alpha + \sigma W \tag{17}$$

where $\alpha = -\ln \lambda$ ($\lambda$ is the location parameter of the generalized gamma distribution) and $\sigma = 1/p$ ($p$ is the scale parameter of the generalized gamma distribution). The density function of $W$ can be represented by the following normalized function

$$f(w, k) = \frac{\exp(kw - e^w)}{\Gamma(k)} \tag{18}$$

where $k$ is the control shape parameter. When $k = 1$ we have the ordinary (non-generalized) extreme-value distribution. The following relationships between the GEV and the GG distributions can be derived:

If $Y \sim GG(\lambda, p, k)$ then $\ln Y \sim GEV(-\ln \lambda, 1/p, k)$

If $Y \sim GEV(\alpha, \sigma, k)$ then $(Y - \alpha)/\sigma \sim GEV(k)$ and $\xi = 0$

**Remark 7.** Without lack of generality, we use in our context the GG family for $T$ but we could also use the generalized $F$ distribution to which nesting family the GG belongs.

## 3.4 Survival Models with Covariates

The models described so far are univariate models designed to provide an adequate framework to fit duration data where the relevant parameters are solely those that characterize the underlying probability distribution function, i.e. the location, the scale and the shape. However, there are good reasons to think that duration data depend on several different exogenous factors. For example, the time from diagnosis of a disease until death may vary according to the personal characteristics of the patient, the type of treatment used, and so on. The length of time between stock market transactions may depend on the characteristics of the stock exchange, the strength and size of the firm, market liquidity, general economic conditions, etc.

There are several approaches that can be used in order to accommodate the dependence of duration data on exogenous factors. We shall consider 3 cases: 1) simple parametric modeling, 2) accelerated failure time and 3) proportional hazards (Menezes, 2004).

### 3.4.1 Simple Parametric Modeling

A general approach to model duration data on a set of exogenous variables is to let the distribution parameters (or their logs) depend in a linear way on a set of relevant covariates $X$.

**Definition 10.** Let $X$ be a vector of covariates and $\beta_\theta$ the underlying vector of parameters. $\theta$ denotes a set of parameters that describe the distribution function of $T$ whose natural logarithm is a linear combination of the covariates included in $X$, such that

$$\ln \theta = X'\beta_\theta \tag{19}$$

If $T$ has a generalized gamma distribution then $\theta = \{\lambda, p, k\}$. We then replace each specific parameter in the above expression in order to obtain separate relationships. For example, $\ln \lambda = X'\beta_\lambda$ and $\ln p = X'\beta_p$. In each case a distinct vector of covariate parameters $\beta_\theta$ is obtained. Each distribution parameter is modeled in an independent way, i.e. holding fixed the remaining distribution parameters. As a general case we may use a function of $\ln \theta$ rather than the logarithm itself. In this circumstance we use $f(\ln \theta) = X'\beta_\theta$, where $f(.)$ may differ for each parameter of the distribution. The prime symbol denotes the transposed matrix.

### 3.4.2 Accelerated Failure Time

A better approach to model duration data as a function of a set of covariates is to expand the linear model considered above and take full advantage of the relationship between the GEV and the GG, as described earlier.

**Definition 11.** Let $W$ be a random variable that follows a generalized extreme-value distribution and $X$ a vector of independent covariates, such that

$$Z = \ln T = -X'\beta + \sigma W \tag{20}$$

where $\{\beta, \sigma\}$ are parameters to be estimated ($\sigma$ denotes the standard deviation of $W$, which is assumed to be constant) and $Z$ has the same distribution of $W$. This means that for every nested distribution of $W$, (extreme value, normal, etc.) there is a corresponding nested distribution for $T$ (Weibull, log-normal, etc.). Additionally, we can consider $\epsilon = \sigma W$, where $\epsilon$ denotes the random error term or disturbance of $Z$.

**Remark 8.** The multivariate equation (20) is closely related to the univariate equation (17). For example, if $W$ has an extreme-value distribution then $T$ has a Weibull distribution, where $\ln \lambda = X'\beta$ and $p = 1/\sigma$. Thus, while $\lambda$ is a function of the covariates, $p$ remains fixed if the error term $\epsilon$ is homoskedastic.

This model is known under the name of Accelerated Life Time or Accelerated Failure Time (AFT) model. We shall use the second designation in this paper. The term "accelerated" is used because the covariates act as multipliers on the survival time. Suppose that $X$ only contains one exogenous variable denoting the age of an individual. The underlying $\beta$ coefficient provides the required information for measuring the effect of a one year change in the age on the baseline survival time. Note that the above multivariate model can be re-specified as $T \exp(X'\beta) = \exp(\sigma W)$. Note also that some covariates may be time-varying, in which case one should replace $X$ by $X(t)$.

Given the multiplicative effect of the covariates on the survival time, it is important to define the baseline state, here denoted by $T_0$. In longitudinal duration models, $T_0$ refers to the initial duration, or initial state, which occurs when $X = 0$. Therefore, $T_0 = \exp(\sigma W)$ and $T$ is distributed as $T_0 \exp(-X'\beta)$.

**Definition 12.** The baseline survivor function of the random variable $T$ is given by

$$S_0(t) = P(T_0 > t) = P\left(W > \frac{\ln t}{\sigma}\right) \tag{21}$$

Note that the baseline survivor function is the probability that the reference subject lasts beyond time $t$.

**Definition 13.** The survivor function of $T$, given $X = x$, will be then represented by

$$S(t,x) = P(T > t \mid X = x) = P\left(T_0 e^{-x'\beta} > t\right) = P\left(T_0 > t e^{x'\beta}\right) = S_0\left(t e^{x'\beta}\right) \quad (22)$$

where the multiplier of the baseline survival time is $\exp(x'\beta) \in \mathbb{R}^+$. The interpretation of this multiplier is that, for example, if $\exp(x'\beta) > 1$, the duration of time in reference (time to death, crisis persistency, etc.) speeds up for subjects having the attribute $X$, relatively to the baseline, by a factor equal to the value of the multiplier. Otherwise, if $0 < \exp(x'\beta) < 1$, the effect is attenuated. This is equivalent to say that the effect quickens for $x'\beta > 0$ and reduces for $x'\beta < 0$. As an analogy, one may think about the number of (Earth) calendar days that Mercury takes to complete an orbit around the Sun by comparison with, e.g. Neptune.

Besides the multiplicative effect of the covariates, the baseline survival time also depends on the shape of the distribution function that describes $W$ and, consequently, $T$. This effect may be time-invariant or time-varying. If it is time-varying, the effect may be monotonic or non-monotonic. Each particular case gives rise to a specific distribution of the random variables $T$ and $W$, as described earlier when we talked about duration dependence.

**Definition 14.** For the random variable $T$ that follows a generalized gamma distribution, given $X = x$, we define the density function and the hazard function as

$$f(t,x) = f_0\left(t e^{x'\beta}\right) e^{x'\beta} \quad (23)$$

$$\lambda(t,x) = \lambda_0\left(t e^{x'\beta}\right) e^{x'\beta} \quad (24)$$

where $\exp(x'\beta)$ denotes the multiplier factor.

**Remark 9.** It is important to note that interpretation of the multiplier effect differs between the survivor and the hazard functions. In terms of survival, the probability that the subject with attributes $X$ survives at any time $t$ is the same as the probability that the reference subject would survive at time $t \exp(x'\beta)$. In terms of hazard, the multiplier means that the underlying subject is exposed at any time $t$ to $\alpha$ times the risk of the reference subject with $\alpha$ times the surviving time, with $\alpha = \exp(x'\beta)$. See, e,g. Kalbfleisch and Prentice (2002).

**Remark 10.** One problem that may arise with the estimation of AFT models occurs when the multiplier yields large effects. Stretching the time axis allows for comparison of different survivor functions but in some cases it may result into distributions belonging to rather different families, turning difficult the use of the extreme-value theory.

### 3.4.3 Proportional Hazards

The proportional hazards model offers an alternative way to model duration data without needing to specify *a priori* the shape of the distribution function. In this sense, it may be considered a generalization of the parametric models described in the previous section (Cox, 1972).

**Definition 15.** If $T$ is a random variable that has an unspecified distribution function and $X$ is a vector of independent covariates, the survivor function under the proportional hazards (PH) model is given by

$$S(t,x) = \{S_0(t)\}^{\varphi(x,\beta)} = \{S_0(t)\}^{\exp(x'\beta)} \tag{25}$$

where the baseline survivor $S_0(t)$ is raised to a power that, in our case, is an exponential function of the covariates. However, if appropriate, one could use another specification for $\varphi(x,\beta)$.

The interpretation in terms of survival is that the probability that the subject with attributes $X$ survives at any time $t$ is the same as the probability that the reference subject would survive at time $t$ raised to a power $\alpha = \varphi(x,\beta)$, where the underlying subject is exposed to $\alpha$ times the risk of the reference subject at any time $t$.

**Definition 16.** If $T$ is a random variable that has an unspecified distribution function and $X$ is a vector of independent covariates, the underlying hazard function is given by

$$\lambda(t,x) = \lambda_0(t)\varphi(x,\beta) = \lambda_0(t)e^{x'\beta} \tag{26}$$

where $\lambda_0(t)$ denotes the baseline hazard and $\varphi(x,\beta) = \exp(x'\beta)$ is the relative risk associated with $X$. That is, the baseline hazard is the hazard for the reference subject which corresponds to $\varphi(x,\beta) = 1$.

Now, the interpretation of the hazard is that the covariates $X$ lead to an increase or decrease of the hazard, relatively to the reference subject, by a proportion $\alpha$ at any duration time $t$. The baseline hazard $\lambda_0(t)$ can be now specified as a parametric function of $T$, e.g. Weibull, gamma, etc., which is independent of $X$.

**Remark 11.** The logarithmic derivative of $\lambda(t, x)$ relative to $x$ is $\beta$. Thus, the vector of covariate parameters can be interpreted as the constant proportional effect of $X$ on the conditional probability of ending a time spell.

Kalbfleisch and Prentice (2002) show that proportional hazards and accelerated failure time models only lead to similar estimates for the Weibull family (which of course includes the exponential if $p = 1$), i.e. the Weibull distribution is the only family that is closed under both proportional hazards and accelerated failure time models. In order to see this, consider the $n$-dimensional covariate vector $x_j = \{(-\ln t/\beta_i^*)I_j\}$, where $I_j = 1$ if $j = i$, and 0 otherwise and $i, j = 1, \cdots, n$. The proportional hazards and accelerated failure time models yield the same results if the underlying hazard functions are equal, that is

$$\lambda_0(t)e^{x'\beta_i} = \lambda_0^*(te^{x'\beta_i^*})e^{x'\beta_i^*} \tag{27}$$

where $\lambda_0(.)$ Is the baseline hazard of the AFT model and $\lambda_0^*(.)$ Is the baseline hazard of the PH model. Replacing $x_i = -\ln t/\beta_i^*$ when $j = i$ we obtain the following equality

$$\lambda_0(t)e^{-(\ln t)\beta_i/\beta_i^*} = \lambda_0(te^{-\ln t})e^{-\ln t} \tag{28}$$

where, under the hypothesis of equal hazards, when $x = 0$ we have $\lambda_0(t) = \lambda_0^*(t)$. The previous equality leads to

$$\lambda_0(t) = \lambda_0(1)t^{\beta_i/\beta_i^*-1} \tag{29}$$

If the condition being tested is true then $\beta_i/\beta_i^* = p$ for every $i$ and thus $\lambda_0(t) = \lambda_0(1)t^{p-1} = \lambda^p p t^{p-1}$, which denotes the hazard function for the Weibull distribution (when $p = 1$ we have the exponential hazard function $\lambda_0(t) = \lambda$). The covariate parameters $\beta^*$ and $\beta$ are proportional to each other, where $p$ is the proportionality constant (they are of course equal for the exponential distribution).

In PH models the effect of covariates is multiplicative in relation to the hazard, whereas in the AFT models, this effect is multiplicative in relation to the survival time (Allison, 2004; Box-Steffensmeier and Jones, 2004; Cox and Oakes, 1984; Kleinbaum and Klein, 2005).

## 3.5   Likelihood Function

The models discussed so far can be empirically estimated by maximum-likelihood, where the function to be maximized is the product of the individual observation likelihoods for the sample under analysis.

**Definition 17.** Let $T$ be a non-negative continuous random variable representing the time until some event of interest, with density function $f(t)$, hazard function $\lambda(t)$ and survivor function $S(t)$. Let $n$ be the size of a sample of subjects for whom duration time occurrences were recorded. We know the pair $\{t_i, d_i\}$ denoting, respectively, the survival time and a censoring dummy for each subject $i$. The censoring dummy $d_i$ is the event indicator or survival status for the $i^{\text{th}}$ subject ($d_i = 1$ if the event has occurred and $d_i = 0$ if the spell duration is still ongoing at the end of the study, i.e. it is right censored). We also know a vector of covariates or risk factors for the $i^{\text{th}}$ subject, denoted by $x_i$. Given this information set, the contribution to the likelihood made by a spell still in progress at the end of the study is (Hosmer and Lemeshow, 1999; Klein and Moeschberger, 2003; Le, 1997) present an analytical treatment of other forms of censoring).

$$P(T, d_i = 0) = P(T > t \mid d_i = 0) \times P(d_i = 0) = S(t^*) \tag{30}$$

and the contribution to the likelihood made by a completed spell is

$$P(T, d_i = 1) = P(t \leq T < t + \Delta t \mid T \leq t^*) \times P(T \leq t^*) = f(t) \tag{31}$$

where $t$ denotes the duration of a completed spell and $t^*$ denotes the duration of a spell still ongoing. Combining these two sets of information we obtain the following likelihood function, assuming independent observations

$$L(\theta) = \prod_{i=1}^{n} f(t_i|x_i)^{d_i} S(t_i|x_i)^{1-d_i} = \prod_{i=1}^{n} \lambda(t_i|x_i)^{d_i} \, e^{-\Lambda(t_i|x_i)} \tag{32}$$

where $\theta$ denotes a vector of parameters to be estimated. Our objective is to estimate the conditional probability of ending a spell duration at time $t + \Delta t$ ($\Delta t \to 0$) given that the spell lasted up to $t$. In practice, the function to be maximized is the log-likelihood, i.e. $\ln L(\theta)$. The maximization of the likelihood function can be numerically done using a procedure such as the Newton-Raphson.

**Definition 18.** The log-likelihood function to be maximized can be written as

$$\ln L(\theta) = \sum_{i=1}^{n} d_i \ln[\lambda(t_i|x_i)] - \sum_{i=1}^{n} \Lambda(t_i|x_i) \tag{33}$$

where $\lambda(t_i|x_i) = \lambda_0 \left( t_i e^{x_i'\beta} \right) e^{x_i'\beta}$ and $\Lambda(t_i|x_i) = \Lambda_0 \left( t_i e^{x_i'\beta} \right) t_i e^{x_i'\beta}$. So, the log-likelihood function can be written as

$$\ln L(\theta) = \sum_{i=1}^{n} d_i \ln \left[ \lambda_0 \left( t_i e^{x_i'\beta} \right) e^{x_i'\beta} \right] - \sum_{i=1}^{n} \Lambda_0 \left( t_i e^{x_i'\beta} \right) t_i e^{x_i'\beta} \tag{34}$$

If the baseline hazard is specified as an exponential distribution where $\lambda_0(u_i) = 1$, we obtain the following special case where $u_i = t_i \exp(x_i'\beta)$

$$\ln L(\theta) = \sum_{i=1}^{n} d_i x_i'\beta - \sum_{i=1}^{n} t_i e^{x_i'\beta} \tag{35}$$

Differentiating $\ln L$ with respect to $\theta$ we obtain the so-called score vector. Equating this vector to zero we obtain the first order conditions for the existence of an optimum (Bagdonavicius and Nikulin, 2002):

$$u(\theta) = \frac{\partial \ln L}{\partial \beta} = \sum_{i=1}^{n} d_i x_i' - \sum_{i=1}^{n} t_i x_i' e^{x_i'\beta} = 0 \tag{36}$$

The Hessian matrix for the log-likelihood function is given by

$$H(\theta) = \frac{\partial^2 \ln L}{\partial \beta \, \partial \beta'} = - \sum_{i=1}^{n} t_i x_i x_i' e^{x_i'\beta} \tag{37}$$

It can be shown that $H(\theta)$ is always definite negative (Kiefer, 1988), leading to a maximum.

The likelihood function as defined above can be used to estimate both AFT and PH models. The estimation by maximum-likelihood implies the specification of the shape of the baseline hazard function.

**Definition 19.** Cox (1975) proposed the estimation of the proportional hazards model using the partial likelihood method which maximizes the following function

$$PL(\theta) = \prod_{i=1}^{m} \frac{\lambda_0(t)\varphi(x_i, \beta)}{\sum_{j=i}^{n} \lambda_0(t)\varphi(x_j, \beta)} = \prod_{i=1}^{m} \frac{e^{x_i'\beta}}{\sum_{j=i}^{n} e^{x_j'\beta}} \tag{38}$$

The partial likelihood measures the weight of the hazard for subject $i$ that is completing its spell duration at time $t$ on the sum of the hazards for all individuals that are at risk at that time. $m$ denotes the number of completed spells. Note that the computation of the partial likelihood does not depend on the specification of any particular baseline hazard function. Only the $\beta$ parameters are estimated by PL and this is why the proportional hazards is a semi-parametric model. The PL function also assumes that spell's duration are independent of each other and, therefore, there are no ties between spells. Partial likelihood can be also maximized using Newton-Raphson method.

**Definition 20.** The partial log-likelihood function can be written as

$$\ln PL(\theta) = \sum_{i=1}^{m} \left( x_i'\beta - \ln \sum_{j=i}^{n} e^{x_j'\beta} \right) \tag{39}$$

The maximization starts with an initial guess at the point which maximizes the log-likelihood function (Cox and Oakes 1984; Klein and Moeschberger, 2003). After $m$ iterations, the estimate of the point which maximizes $\ln PL(\theta)$ is given by

$$x_{m+1} = x_m - H(x_m)^{-1} u(x_m) \tag{40}$$

where $x$ is a vector of covariates, $u$ is the score vector and $H$ is the Hessian matrix.

## 3.6 Caveats

Having established the basis for the estimation of survival models, there is a number of caveats that should be reminded given the potential loss of efficiency they may cause in the estimation process. The first one refers to recurrent spells. If consecutive spells have some kind of link, we may expect autocorrelation in the error term of the regression model. Otherwise, subjects can be treated as independent observations and $E(\epsilon_i, \epsilon_j) = 0$ $(i \neq j)$.

The second caveat is the possibility that covariates may be time-dependent. A thorough study of the path behavior of covariates over spells with duration greater than 1 will help to design the appropriate time-dependency function.

The third caveat concerns omitted heterogeneity, where unobserved attributes may affect the probability of ending a spell that is otherwise attributed to duration dependence, thus biasing estimated hazards towards negative duration dependence. Omitted heterogeneity is also known as frailty (Gutierrez, 2002; Hougaard, 1995; Wienke, 2003) and there are a number of methods to correct unobserved data (Duchateau and Janssen, 2007; Gjessing *et al*, 2003; Glidden, 1998). So, in our study, we will deal with these problems by using robust standard errors (Freedman, 2006; King and Roberts, 2015; Yamano, 2009) and frailty correction in our maximum likelihood estimates (Chamberlain, 1984; Heckman and Singer, 1984).[4]

---

[4] Note that in our study we don't have problems either with non-informative censored spells (Allison, 2004: 371) or with Gaussian distributional assumptions of the residuals Cleves *et al.* (2004).

# 4   Data Characterization

## 4.1   Economic Recessions and Financial Market Breakdowns

Asset price fluctuations is a recurrent phenomenon in financial markets. Figure 1 displays the monthly evolution of the S&P500 real stock market price index between 1871 and 2016. The data used in this study pertain to the Shiller database.

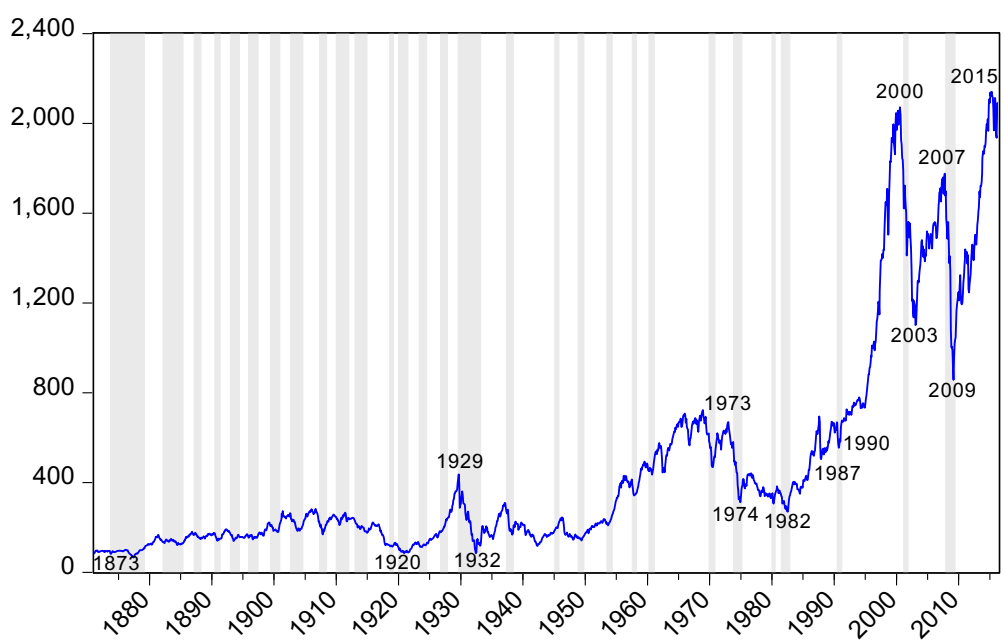

Figure 1. Monthly evolution of the S&P500 real price index, 1871-2016 (1747 obs.). Shaded bars represent US recessions. Source: Shiller's home page

As may be seen, over the last 145 years there were several periods of economic expansion and contraction in the United States.[5] There are noticeably two distinct periods: the first one, going up to the end of the Great Depression and the second from the Great Depression onwards.[6] In the first period there were frequent and lengthy economic recession spells with relatively short intervals between them. Between 1871 and 1938 there were 17 spells of economic recession, i.e. an average of 2.5 per decade. The mean duration of each recession spell was around 22.7 months, ranging from a minimum of 8 to a maximum of 66 months. The average time interval between recessions was 25 months. In terms of financial markets, there were during this period 163 spells of stock market price decline, i.e. an average of 2.41 per year.

---

[5] US recession spells in Figure 1 are signaled by shaded bars.

[6] We include in the first period the pre-World War II recession of 1937-38, because it is often seen as yet a result of the Great Depression.

In many cases, spells of economic recession occur together with spells of financial crisis, usually following them. Just to mention a few, this is the case of the panic of 9 May 1873 that initiated the Long Depression in the United States and much of Europe and lasted more than five years. The post-World War I recession of 1918-19 and the depression of 1920-21, spanned more than two years during which a falling-off in stock market prices occurred. The effects of the Great Wall Street Crash of 24 October 1929 lasted over four years, leading to the Great Depression which was followed by short replicas such as the recession of 1937-38. Nevertheless, not all stock market price falls are associated with specific economic recessions. For example, in 1916-17 there was a spell of stock market price decline lasting four months and another one lasting six months without evidence of economic recession in the US during this period.

After the Great Depression, spells of economic recession became scarcer and shorter but stock market breakdowns appear to be slightly more frequent. Between 1938 and 2016 there were just 12 spells of economic recession in the US, i.e. an average of 1.5 per decade. The mean duration of each recession spell was around 11.8 months, ranging from a minimum of 7 to a maximum of 19 months. The average time interval between recessions was 61 months. In terms of financial markets, there were during this period 199 spells of stock market price decline, i.e. an average of 2.55 per year.

The post-World War II period was in general prosperous and tranquil for almost thirty years (1945-73), both in terms of economy and of financial markets. In January 1973 there was a stock market crash in the UK, whose effects lasted until the end of 1974, due to the oil crisis and the miners' strike. The UK stock market crash spread over many other stock markets worldwide, including the US. In August 1982 there was the Souk Al-Manakh (Kuwait) stock market crash with widespread contagion to the US. The Black Monday occurred in 19 October 1987. Following the Kuwait invasion by Iraq in July 1990, there was a sharp drop of the Dow Jones Industrial Average (DJIA), known as the Early 1990s crisis. In the outset of the new millennium there was a series of events that disturbed the stock markets across the US. The collapse of a technology bubble in 10 March 2000 triggered the Dot-com crisis. The September 11 attacks in 2001 had serious economic effects, causing global stock markets to drop sharply. The stock market downturn of 9 October 2002 affected stock exchanges across the world, leading to lows last reached in the late 1990s. Like before the Great Depression, many stock market break-downs occurred during periods not classified as economic recession.

The next major event was the US bear market of 2007-09, started in 11 October 2007. The DJIA, Nasdaq Composite and S&P500 all experienced sharp drops from their peaks in 2007. In September 2008, the US subprime crisis quickly devolved into a global crisis resulting in a number of bank failures in Europe and sharp falls in the value of equities and commodities worldwide. Finally, the 2015-16 Chinese stock market crash started in 12 June 2015 and lasted until January 2016.

Overall, the stock market price data in Figure 1 reveal a few large movements (up and down) along with other intermediate fluctuations. Before World War II, from the low of 1920 (84.6) to the high of 1929 (436.0), real stock market prices rose five times. This was lost in three years and in 1932 (84.5) real prices shrunk to the 1920's level. Six years later real prices doubled to 174.5 but still remained less than half-way of the peak of 1929. The next major cycle lasted more than three decades, ending in 1973 (669.7). During this period real prices rose consist-ently almost four times. However, in just two years, real prices decreased more than 50% and in the end of 1974 dropped to 311.4, a value last observed only in the 1950s. The next cycle started in 1982 (270.4) and real prices boosted more than seven times until 2000 (2071.5). Another break of almost 50% ended three years later, in 2003 (1101.6). Real prices rose another time between 2003 and 2007 (1775.7) and dropped again 50% until 2009 (857.7), a value last observed only in the mid-1990s. Since then, real prices increased again two and a half times until 2015 (2140.0), completing the $\mathcal{W}$-shaped crisis initiated in 2000.[7] The value observed in May 2015 is the highest real price ever recorded in the S&P500 since 1871.

## 4.2 Stock Market Returns

Figure 2 depicts the monthly evolution of the S&P500 stock market returns between 1871 and 2016. Stock market returns are computed as the difference between the logarithm of stock market prices.

---

[7] By $\mathcal{W}$-shaped crisis (or double-dip crisis) we mean a period where real prices fall sharply, partially recover for a short period and then fall back to the previous low before recovering again to the previous peak, behaving in a pattern resembling a $\mathcal{W}$ (National Bureau of Economic Research - NBER). This happened with the S&P500 be-tween 2000 and 2015.

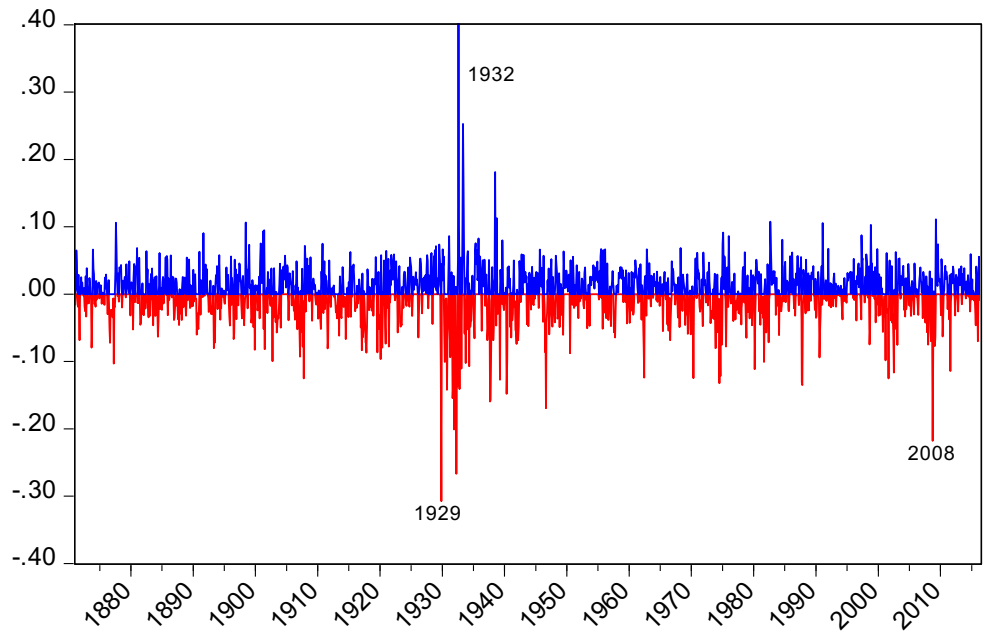

Figure 2. Monthly evolution of the S&P500 stock market returns, 1871-2016 (1746 obs.). Positive returns are in blue and negative returns are in red

The data clearly reveal two periods of greater instability in stock market returns. The first one, and by far the most prominent, was the Wall Street Crash of 1929 with an average monthly return of -0.03 and a high-low range of $-0.31$ (Nov/1929) to 0.41 (Aug/1932). The second period was the US bear market of 2007-09 with an average return of -0.02 and a high-low range of $-0.22$ (Oct/2008) to 0.11 (Apr/2009). Excluding outliers the high-low range is trimmed between $-0.08$ and 0.09. There are 67 outliers (extreme values) in the sample, representing 3.8% of the total. More than 20% of these extreme values occurred during the two crashes mentioned above.[8]

## 4.3  Asset Price Declines

Although abnormal asset price fluctuations may include both positive and negative returns, we are interested in analyzing the behavior of negative returns given their implications in terms of market instability and their likely connection with economic recessions. It is well known that higher volatility in stock market returns is associated with periods of falling prices (Boubakri *et al.*, 2016; Li and Liaw, 2015). On the other hand, when prices fall more than a certain level (usually 20%) for at least two consecutive months and investors become too distrustful regarding their prospective gains, causing a negative sentiment in the market, then the market enters into a bear phase (Frugier, 2016; Ntantamis and Zhou, 2015; Wu and Lee, 2015). Bear markets based on the S&P500 index occurred several times over the last 145 years, namely during the Long Depression, the post-World War I recession and ensuing depression, the Wall Street crash and Great Depression, the 1974 oil crisis, the Black Monday, the 9/11 attacks, the subprime crisis, etc. Bear markets are also related to volatility clustering and persistence.

---

[8] The size of the high-low range during the bear market of 2007-09 (0.33) is less than half the same size during the Wall Street crash. (0.72). Furthermore, the former is almost twice the size of the high-low range after dropping outliers or extreme events (0.17).

Just to give some examples, during the Great Depression there were several episodes of stock market fall-downs lasting more than two months with losses greater than 30% in each of them; on the other hand, the US crash of 2008 carried out losses of almost 30% during a relatively long period. Therefore, bear markets are intrinsically related to the persistence of high volatility and should not be confused with market price corrections that usually are short-termed. One key question is then what is the probability of ending a period of stock market price decline? And how does this probability evolve over the duration of such period?

Between 1871 and 2016, in the US S&P500 there were 761 monthly episodes of market price decline (negative returns), of which 270 (35%) were joint episodes of declining prices and US recession. A total of 362 spells (2.49 per year) of negative returns were recorded where a spell is defined as a period of consecutive monthly episodes wherein the market price declines. The length of a spell is denoted by duration or spell duration. The average spell duration of negative returns within the sample period analyzed is 2.10 months and the median is 2 months. The minimum duration of a spell is one month and the maximum is twelve months. Additionally, the mean loss per month during spells of negative returns is 2.63%, reaching a maximum of 18.21% in December 1931. In general, there is a tendency for increasing monthly price declines as the spell duration become longer.

### 4.3.1  Spell Duration of Asset Price Decline

As noted above, a total of 362 events of asset price decline were observed in the S&P500 index between 1871 and 2016. An event occurs when a sequence of episodes of the relevant phenomenon ends and the spell duration is completed. In our study, the sequence of episodes refers to a succession of consecutive monthly observations of negative returns and the event occurs when this sequence is interrupted by an episode of non-negative returns. Our event of interest is, therefore, a completed spell duration of declining asset prices and we have records of spells lasting from one till twelve months. The distribution of spell's duration is represented in Figure 3.

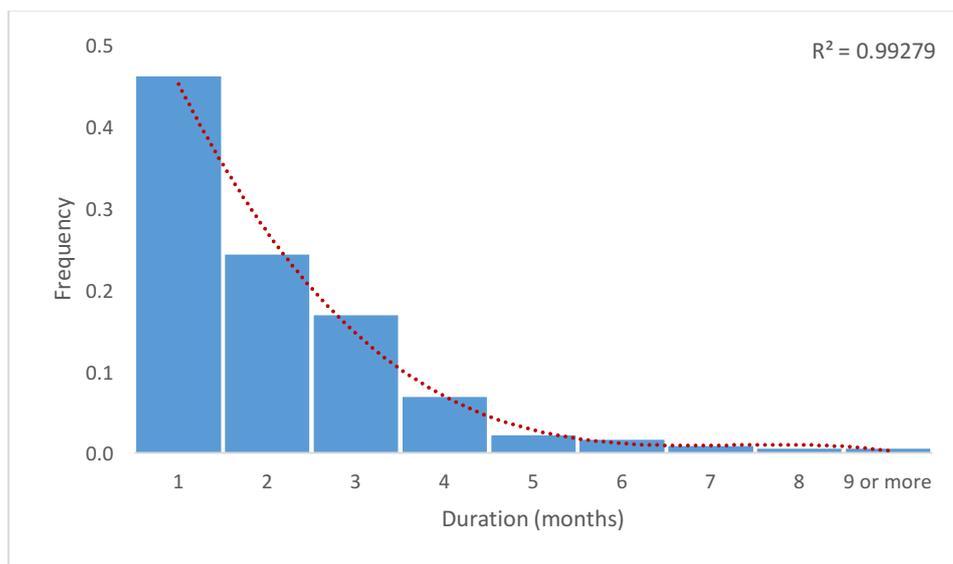

Figure 3. Histogram of the duration time of price declines for the S&P500, 1871-2016 (362 spells). The dotted red curve represents a third-degree polynomial trendline fit $\hat{y} = -0.0018t^3 + 0.0398t^2 - 0.288t + 0.7019$ where $t$ denotes the duration of the spell

Some 46% of the total number of spells lasted just one month. About 24% lasted two months, 17% lasted three months and 7% lasted four months. The remaining 6% lasted five months or more. The distribution is skewed and has a long right-tail. A third-degree polynomial 'growth' model $p(t) = \sum_{k=0}^{3} \gamma_k t^k$ fits the data with a goodness-of-fit coefficient of 0.99. Predicted probabilities depart from observed frequencies with a squared error of less than 0.1% in all cases except the uppermost duration value (12 months).

## 4.4 Covariates

The mean duration of completed spells between 1871 and 1938 is 2.20 months with variance of 2.84 and maximum of 12. On the other hand, the mean duration of completed spells between 1938 and 2016 is 2.03 months with variance of 1.72 and maximum of 8.[9] This leaves a mean difference of 0.17 and a *t*-test (*t* = 1.06; df = 302) for equal means and known variances does not reject the null hypothesis (two-tail *p*-value = 0.29). We may then conclude that spells duration of stock market decline before and after the end of the Great Depression do not differ significantly in statistical terms, and thus, data segmentation on the basis of this (calendar time) criterion is not an adequate choice.

A different picture is obtained when we compare the mean duration of completed spells during periods of economic recession and periods of economic boom. In fact, during recessions the mean duration of completed spells is 2.60 with variance of 3.85 and maximum of 12, whereas for booms the mean duration of completed spells is 1.88 with variance of 1.32 and maximum of 7. The mean difference reaches 0.72 and the *t*-test (*t* = 3.65; df = 150) does reject the null of equal means (two-tail *p*-value < 0.01). Therefore, it makes sense to use a recession/no recession dummy variable to compare results of the survival models under different economic environments.

Similarly, the mean duration of completed spells differs significantly with the degree of intensity of the monthly price decline, in particular in what concerns the tails of the distribution. The mean difference between the upper and the lower quartiles of the price decline distribution is 1.04 which is significantly different from zero at less than 1% (*t* = 6.04; df = 119). The mean duration in the lower price decline quartile also differs significantly at the 1% level from the mean duration in the interquartile range. Thus, the intensity of monthly price decline will also be used as a continuous regressor in our empirical model.

Other regressors such as the inflation rate, dividends and earnings were also attempted, but no significant impact on spell duration was observed and the regressors were dropped. Some results of the *t*-tests carried out in this context are presented in Table 1. The results clearly show that the mean duration time of price declines varies significantly with the intensity of the price variation and the general economic condition. We shall also add a control variable for the long-term interest rate but the 145-years dataset used in this study does not contain other control variables affecting spell durations.

---

[9] The minimum is 1 in both cases.

|  | World War II | | Economic Recessions | | Price Decline | | | |
|---|---|---|---|---|---|---|---|---|
|  | Before 1938 | After 1938 | Recession | No Recession | Q1 | Q2 | Q3 | Q4 |
| Mean | 2.19632 | 2.02513 | 2.59649 | 1.87500 | 1.37363 | 2.13333 | 2.51111 | 2.39560 |
| Variance | 2.83776 | 1.72159 | 3.85344 | 1.32439 | 0.39219 | 2.45393 | 3.08414 | 2.26398 |
| Observations | 163 | 199 | 114 | 248 | 91 | 90 | 90 | 91 |
| df |  | 302 |  | 150 |  | 116 | 111 | 120 |
| $t$-stat |  | 1.06046 |  | 3.64676 |  | -4.27533 | -5.79129 | -5.98184 |
| P($T <= t$) two-tail |  | 0.28978 |  | 0.00037 |  | 3.94E-05 | 6.6E-08 | 2.34E-08 |
| $t$ Critical two-tail |  | 1.96785 |  | 1.97591 |  | 1.98063 | 1.98157 | 1.97993 |

Table 1. Two-sample $t$-tests assuming unequal variances for the mean duration time of price declines for the S&P500, 1871-2016 (362 spells). H$_0$: $\mu_1 - \mu_2 = 0$

## 4.5   Survival Time and Failure Rates of Price Declines

A preliminary analysis based on the Kaplan-Meier (1958) estimator reveals that survival probabilities start on 0.50 in periods of no economic recession and on 0.62 in periods of economic recession (Figure 4). Over the entire span the former remain substantially below the latter and die out earlier. Standard errors lie around 0.02 in periods of no recession and 0.03 in periods of recession. On the other hand, in periods of no recession the observed survival rates tend to lap up the predicted ones, but in periods of recession observed survival rates tend to be higher than predicted ones. Therefore, Kaplan-Meier seems to underestimate the survival probabilities especially during periods of recession.

Figure 5 presents the Kaplan-Meier estimates for the survival and hazard rates, comparing now the shapes of the functions for recessions/no recessions. The left panel is basically similar to the graph presented in Figure 4, but now we can see that as spells become longer the survival probabilities tend to approach one another and the biggest difference (0.21) is recorded between 2-3 months.

The right panel shows the smoothed hazard estimates for recessions/no recessions where we can see a totally different pattern for each subsample. In fact, the hazard rate seems to increase over time in periods of economic growth and is substantially higher than the hazard rate estimated for periods of economic recession. As time goes on, the hazard rate in periods of recession tend to decrease and, therefore, depart from the hazard rate in periods of no recession. The overall picture shows a diversity of situations that will be further investigated in the next section.

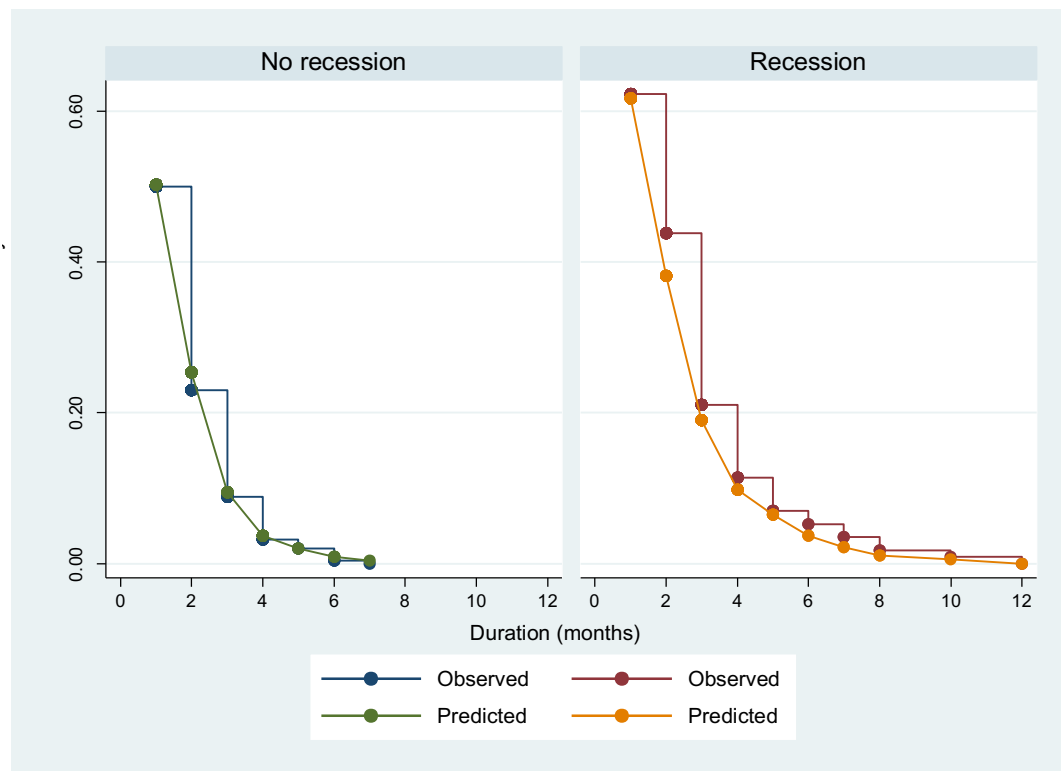

Figure 4. Kaplan-Meier observed and predicted survival of the duration time of price declines for the S&P500, 1871-2016 (362 spells). Recession curves in red (right) and no recession in blue (left)

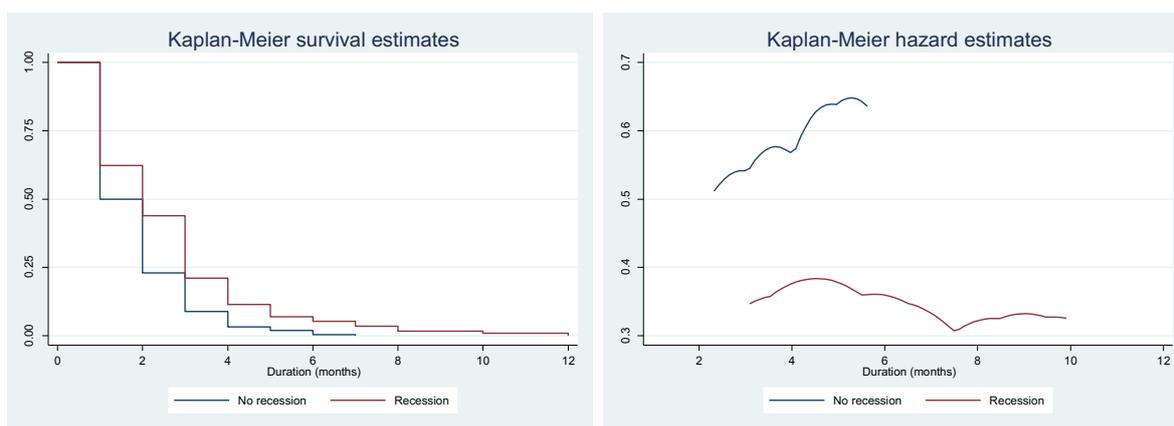

Figure 5. Kaplan-Meier survival (left) and hazard (right) estimates of the duration time of price declines for the S&P500, 1871-2016 (362 spells). Recession curve in red and no recession in blue

# 5   Results

## 5.1   Proportional Hazards vs. Accelerated Failure Time

The results of the estimation of the models described in section 3 are presented in Tables 2 and 3. The estimates were initially computed for several distributions nested in the generalized gamma family as well as for the Cox proportional hazards. Some distributions, such as the exponential and the Weibull, can be used either in the context of proportional hazards (PH) or in the context of accelerated failure time (AFT) models. The estimates are comparable (but not equal) in both cases but the PH specification allows one to obtain estimates of the mean time-invariant hazard ratios while the AFT yields estimates of the regression $\beta$-coefficients (also called acceleration factors, which describe the direct effect of an exposure on survival time).[10] The Cox regression is a semi-parametric PH model; conversely, the log-normal only fits into the AFT family. Recall that one advantage of accelerated failure time models is that they allow for time-varying duration dependence, i.e., hazard rates that change over the spell duration.

The $z$-scores and the underlying $p$-values of the estimated parameters are very similar in the PH and AFT models and are identical for the estimates of the distributional parameters. However, the proportional hazards estimates of the coefficient standard errors are not as efficient as those obtained by the accelerated failure time models, since the latter tend to be sometimes smaller than the former.[11] It is therefore important to assess which assumption fits better our data: the proportional hazards or the accelerated failure time.

A likelihood-ratio test of the proportional hazards assumption yields the statistic $\chi^2(3) = 1.72$ ($p$-value = 0.63) and, therefore, the null of proportionality cannot be strictly rejected. The proportional hazards test is a test of the null hypothesis that the slope coefficient in a generalized linear regression of the scaled Schoenfeld residuals on time is equal to zero (Grambsch and Therneau, 1994). However, since some types of non-proportionality cannot be detected by this likelihood-ratio ($LR$) test the choice of the best model needs to be confirmed by an alternative gauge,[12] because non-rejection of the null of the proportional hazards assumption does not imply the rejection of the accelerated failure time assumption. Indeed, the Weibull model for example has the property that if the PH assumption holds then the AFT assumption also holds, and vice-versa. Likewise, the exponential PH is no more than a re-parameterization of the exponential AFT model. However, the same rule does not apply to the Cox or the log-normal models, which may eventually yield better estimates than the exponential or the Weibull.

---

[10] As noted in section 3, the acceleration factor describes the widening or shrinkage of the survival functions when comparing one group to another and is the ratio of survival times corresponding to any fixed value of $S(t)$. See, e.g. Kleinbaum and Klein (2005).

[11] This is especially so in the case of models with time-varying hazards.

For nested models, such as those derived from the generalized gamma family, the choice criterion can be based on the likelihood ratio test $LR = -2(\ln L_0 - \ln L_1)$, where $L_0$ is the initial likelihood (or the likelihood of the model without covariates) and $L_1$ is the likelihood of the full model (Box-Steffensmeier and Jones, 2004; Cleves *et al.*, 2004; Kleinbaum and Klein, 2005). The $LR$ test is asymptotically equivalent to the Wald test with as many degrees of freedom (df) as the number of parameters to be tested. However, in the presence of robust estimates of the parameter's covariance matrix only the Wald test can be used (Cleves *et al.*, 2004). Nevertheless, neither the $LR$ nor the Wald test can be used to compare non-nested models (Box-Steffensmeier and Jones, 2004; Klein and Moeschberger, 2003; Kleinbaum and Klein, 2005; Kuha, 2004). In this case, one can decide on the basis of information criteria. The most used criteria are perhaps the Akaike Information Criterion (Akaike, 1974) and the Bayesian Information Criterion (Schwarz, 1978). Both criteria penalize the log-likelihood of the model against the number of covariates and specific parameters to be estimated. The Akaike Information Criterion, for example, is given by $AIC = -2 \ln L + 2(k_1 + k_2)$, where $k_1$ is the number of covariates in the model and $k_2$ is the number of model specific distributional parameters. The best fit is the one that minimizes the $AIC$ statistic.

In this study we ground our model selection on the Akaike Information Criterion, although the same conclusions would be reached using the Bayesian Information Criterion. As explained above, information criteria are better than $LR$-type tests under robust estimation of non-nested models, as in our case. For a given distribution function the $AIC$ is equal both in the PH and AFT model specifications (exponential and Weibull), but AFT can use a broader variety of distributions, including non-nested ones. The Weibull model minimizes the $AIC$ when the PH metric is used ($AIC = 661.57$). However, the AFT log-normal model yields overall a better fit ($AIC = 613.56$).[13]

---

[12] For instance, unexpected economic/financial effects, delayed recovery, and other barriers may cause unpredictable variations in the hazard rate over time.

---

[13] Since our data does not contain censored spells, these results are also confirmed by a simple goodness-of-fit test based on the sum of squared residuals. In our case, the SSR for the log-normal without frailty has the lowest value of all models tested (SSR = 114.79). The residual′s mean is very close to zero. Other tests based on Cox-Snell and martingale residuals lead to similar conclusions (Box-Steffensmeier and Jones, 2004; Novák, 2010; Therneau et al., 1990).

| Model | Recession | | | Price Decline | | | Interest Rate | | | AIC |
|---|---|---|---|---|---|---|---|---|---|---|
| | $\beta$-coeff. | Std. Err. | $p$-value | $\beta$-coeff. | Std. Err. | $p$-value | $\beta$-coeff. | Std. Err. | $p$-value | |
| **Frailty** | | | | | | | | | | |
| Cox | -0.295 | 0.121 | 0.015 | -0.018 | 0.040 | 0.658 | -0.028 | 0.023 | 0.216 | 3742.69 |
| Exponential | 0.319 | 0.090 | 0.000 | 0.092 | 0.022 | 0.000 | 0.074 | 0.010 | 0.000 | 862.02 |
| Weibull | 0.277 | 0.084 | 0.001 | 0.047 | 0.020 | 0.020 | 0.025 | 0.013 | 0.048 | 661.57 |
| log-Normal | 0.152 | 0.076 | 0.046 | 0.035 | 0.015 | 0.021 | 0.032 | 0.013 | 0.012 | 613.57 |
| | | | | | | | | | | |
| **No Frailty** | | | | | | | | | | |
| Cox | -0.300 | 0.093 | 0.001 | -0.059 | 0.024 | 0.016 | -0.027 | 0.015 | 0.062 | 3746.83 |
| Exponential | 0.278 | 0.088 | 0.002 | 0.050 | 0.021 | 0.017 | 0.026 | 0.013 | 0.049 | 855.01 |
| Weibull | 0.334 | 0.099 | 0.001 | 0.057 | 0.026 | 0.029 | 0.023 | 0.014 | 0.101 | 711.29 |
| log-Normal | 0.207 | 0.073 | 0.004 | 0.040 | 0.016 | 0.014 | 0.029 | 0.012 | 0.018 | 621.54 |

Table 2. Coefficient estimates, robust standard errors and $p$-values of the models estimated with and without frailty correction. Regressors are the recession dummy, average monthly price decline (%) and long-term interest rate (%). AIC = Akaike Information Criterion, 1871-2016 (362 spells). $H_0$: $\beta = 0$

| Model | ln $p$ | | | ln $\sigma$ | | | ln $\theta$ | | | SSR |
|---|---|---|---|---|---|---|---|---|---|---|
| | coeff. | Std. Err. | $p$-value | coeff. | Std. Err. | $p$-value | coeff. | Std. Err. | $p$-value | |
| **Frailty** | | | | | | | | | | |
| Cox | - | - | - | - | - | - | ln[.041] | 0.046 | 0.021 | 397.19 |
| Exponential | - | - | - | - | - | - | -18.066 | 0.140 | 0.000 | 130.88 |
| Weibull | 1.006 | 0.044 | 0.000 | - | - | - | 0.993 | 0.116 | 0.000 | 116.62 |
| log-Normal | - | - | - | -0.806 | 0.129 | 0.000 | -1.076 | 0.575 | 0.061 | 119.57 |
| **No Frailty** | | | | | | | | | | |
| Cox | - | - | - | - | - | - | - | - | - | 480.26 |
| Exponential | - | - | - | - | - | - | - | - | - | 126.00 |
| Weibull | 0.529 | 0.039 | 0.000 | - | - | - | - | - | - | 148.19 |
| log-Normal | - | - | - | -0.574 | 0.032 | 0.000 | - | - | - | 114.79 |

Table 3. Ancillary distributional parameters, robust standard errors and $p$-values of the models estimated with and without frailty correction. Parameters are $p$ (Weibull inverse scale), $\sigma$ (log-Normal scale) and $\theta$ (frailty). Location is given by $\lambda = \exp(X'\beta)$. SSR = Sum of Squared Residuals, 1871-2016 (362 spells). $H_0: \ln\{p, \sigma, \theta\} = 0$

## 5.2 Estimation

All models were estimated using robust standard errors. Robust estimation should be used in the presence of residual heteroskedasticity, i.e. when the variance of the residuals is not constant over the observations. In our case, robust standard errors were computed using the following modification

$$\text{Var}(\hat{\beta}) = (X'X)^{-1}X'\hat{\Sigma}X(X'X)^{-1} \tag{41}$$

where $\hat{\Sigma} = \text{Diag}\{\hat{\epsilon}_i^2\}$, $(i = 1, \cdots, n)$ and 'Diag' denotes the diagonal matrix of $n^{\text{th}}$ order adjusted for $n/(n-k-1)$ degrees of freedom in small samples.

A gamma and an inverse Gaussian frailty correction was also attempted for the parametric regressions, and the frailty $\ln\theta$ estimate of the variance is significant at less than 1% in the exponential and Weibull distributions but just at the 6.1% significance level in the case of the log-normal. In the latter case, there is no warranty that $\ln\theta$ cannot be zero at standard levels of significance and the 95% confidence interval spans $-2.20$ to $+0.05$. Recall that frailty models are used to account for omitted heterogeneity, in the sense that some subjects are more susceptible to fail than others for unknown or unmeasured reasons (Blossfeld *et al.*, 2007; Box-Steffensmeier and Jones, 2004; Hougaard, 2000; Karim, 2008; Therneau and Grambsch, 2000; Wienke, 2003). Therefore, in our study, robust estimation with frailty correction is used in order to account for heteroscedasticity and unobserved heterogeneity. Although the presence of unobserved heterogeneity is not absolutely clear in our log-normal model, this model also yields the lowest *AIC* value when estimated without any frailty correction (in this case *AIC* = 621.54). The improvement in the fit of the log-normal with frailty against the log-normal without frailty is modest (1.3%) but it becomes five times as big as that (7.3%) when the log-normal fit is compared with the Weibull fit, both with frailty correction. For the parametric models tested in this study, the exponential is the one that yields the worst fit of all the three parameterizations. We shall therefore rely on the results of the accelerated failure time log-normal model.

### 5.2.1 Parameter Estimates

The results discussed in this section refer to the log-normal model with and without frailty correction as presented in Table 2. The $\beta$-coefficients are the acceleration factors for each covariate as well as the constant term. The models include three covariates: *i*) a recession dummy (***Recession***) indicating whether the spell of market price decline occurs during an US economic recession or not, *ii*) a continuous variable denoting the monthly average of the underlying price decline, in percentage (***Price***), and *iii*) a continuous variable denoting the monthly average of the long-term interest rate over the underlying spell (***IR***). In Table 3 we also present the robust standard errors and the *p*-values of the *z*-scores for the significance level of the estimates.

The $\beta$-coefficient of the recession dummy varies between 0.15 (log-normal with frailty) and 0.21 (log-normal without frailty) and is significantly different from zero at the 5% level or better in both models. This means that, *ceteris paribus*, during recessions in the US the time to event accelerates by a factor of $\exp(\beta)$. Since in our study the event of interest is the end of a spell of stock market price decline, the acceleration factor indicates how longer or shorter is the survival time (or duration) compared to the baseline survival. When $\beta$ is positive the multiplier is greater than 1 and the duration time increases relatively to the baseline; conversely, when $\beta$ is negative the multiplier is less than 1 and the duration time decreases relatively to the baseline. Of course, when $\beta$ is zero there is no effect and the regressor is dropped. Replacing $\beta$ by the underlying estimates obtained for the log-normal model, we get an acceleration factor placed between 1.16 and 1.23, that is, other things being equal, spells of financial market dwindling are between 16%-23% longer during periods of economic recessions than during periods of economic growth.

The $\beta$-coefficient of the price decline covariate is also significantly different from zero at the 5% level and varies between 0.035 and 0.040. It means that, on average, a one percentage point drop of the market price index is related to an increase of 3.6%-4.1% in the length of a spell of financial market decay, all other things remaining equal. Finally, the $\beta$-coefficient of the long-term interest rate lies between 0.029 and 0.032 and is, as well, significant at the 5% level. *Ceteris paribus*, an increase of one percentage point in the long-term interest rate is related to spells of price decline lasting between 2.9%-3.2% more time than the baseline survival. The constant term is only significant at standard levels in the log-normal without frailty correction.

Ancillary distributional parameters are reported wherever relevant, in addition to the covariate's parameters. For the log-normal, the relevant parameters are $\lambda$ (location), $\sigma$ (scale) and $k$ (shape). Note that $\lambda = \exp(X'\beta)$, i.e. is a function of the covariates at the mean, $\sigma = 1/p$ and $k \to \infty$. The parameter $\theta$ is the gamma frailty parameter for unobserved heterogeneity. All the parameters are statistically assessed on the basis of robust standard errors in order to compute the $z$-scores and the underlying $p$-values.

At the mean values of the covariates, $\lambda$ spans 1.56-1.74, i.e. the subject or spell with average attributes in the dataset is expected to last about 56%-74% more time than the reference (baseline) subject. The average subject has a mean recession time proportion of 0.315, a mean market price decline of 2.63% and a mean long-term interest rate of 4.62%.

The scale parameter $\sigma$ is positive and is placed between 0.45 and 0.56, indicating that the hazard increases from zero to a maximum which is close to the median and then decreases monotonically to zero as $t \to \infty$. The hazard function is therefore hump-shaped (Kalbfleisch and Prentice, 2002; Klein and Moeschberger, 2003; Lee and Wang, 2003), and the scale parameter indicates how quickly the hazard rate rises to its peak (Box-Steffensmeier and Jones, 2004).

### 5.2.2 Hazard Functions

The models estimated in this study yield in some cases curious shapes of the hazard functions, depending on the parametric or semi-parametric distribution used to obtain the estimates. These functions are presented in Figure 6. The Cox PH model yields a nonlinear hazard function with a shape that resembles the contour of the slope of a mountain with step levels. The exponential is represented by a zero-slope straight line as should be. The Weibull hazard increases monotonically as spell duration progresses removing the hypothesis of duration dependence bias due to omitted heterogeneity. Finally, the log-normal exhibits a hump-shaped hazard function, first increasing to its maximum and then decreasing towards zero.

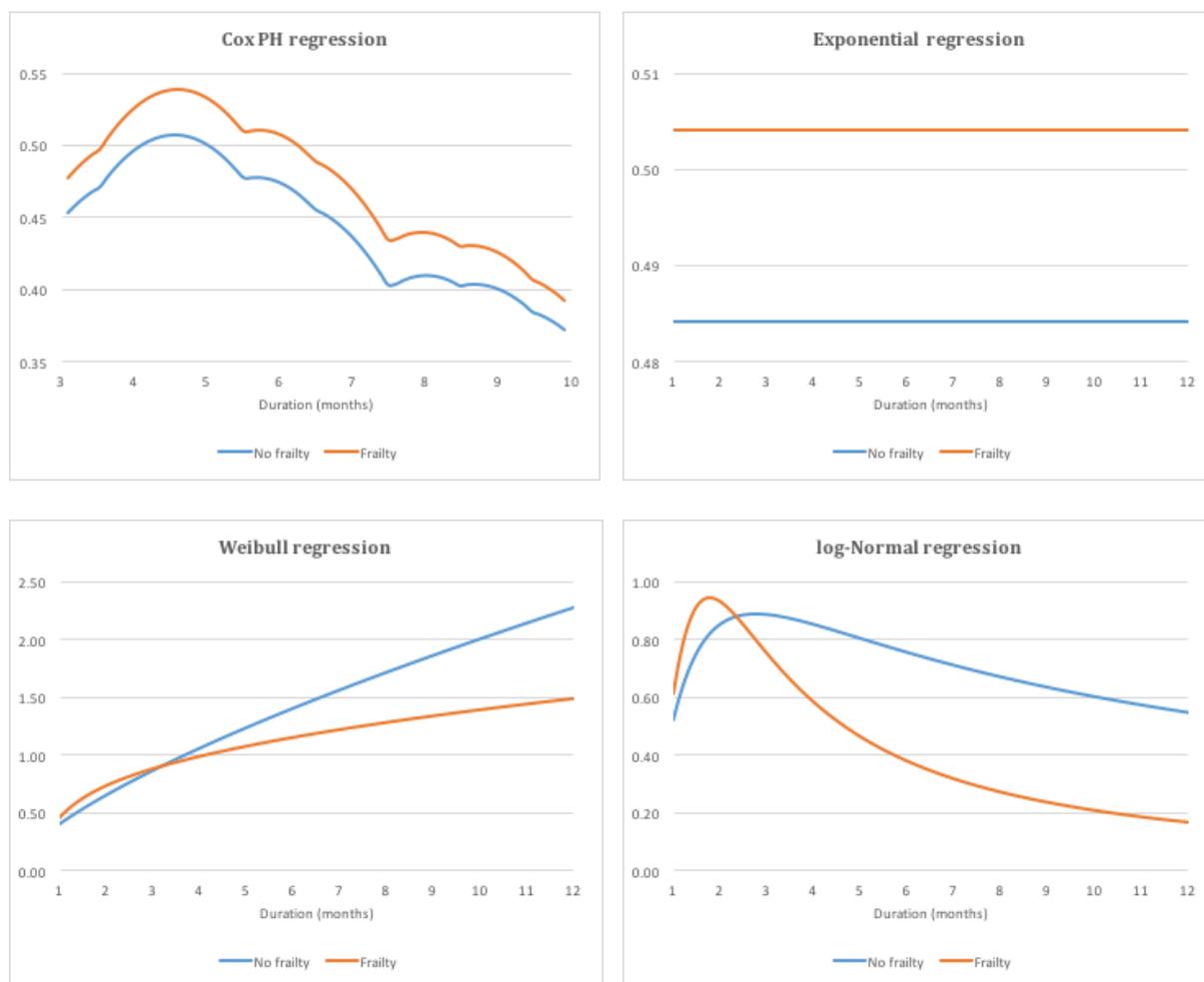

Figure 6. Hazard functions of the duration time of price declines for the S&P500, 1871-2016 (362 spells). Frailty estimates in red and no frailty in blue

The models estimated with frailty correction lead to higher hazards in the Cox and exponential metrics. The mean difference is about 0.03 in the case of the Cox PH model and is 0.02 in the case of the exponential. For the Weibull and the log-normal, the models estimated with frailty correction yield higher hazards than the models estimated without this correction up to a certain level and lower hazards onwards, producing non proportional curves. The turning point occurs at 3.20 months in the case of the Weibull and 2.43 months in the case of the log-normal. The log-normal hazard function with no frailty correction reaches its maximum of 0.89 at the duration of 2.76 months, i.e. about 10 days after the turning point. On the other side, the log-normal hazard function with frailty correction reaches its maximum of 0.94 at the duration of 1.77 months, that is, one month before the maximum of the hazard with no frailty correction and 20 days before the turning point. This is important in order to understand at which point in time duration dependence changes from positive to negative and hysteresis effects become effective. As explained in previous sections, we shall rely on the log-normal results for our conclusions.

The log-normal with no frailty correction begins with a hazard of 0.52 for a spell duration of 1 month and, after reaching its maximum, decays slowly towards a value of 0.54 at 12 months. On the other hand, the log-normal with frailty correction initiates with a hazard of 0.61 for a spell duration of 1 month and, after its maximum, decays quickly to 0.16 at 12 months. Therefore, the model without frailty starts with a lower hazard than the frailty model, rises more (1.71 against 1.54), but takes one month longer to reach the peak, consistently with the values of the scale parameter $\sigma$ discussed above. After twelve months, the hazard remains approximately at the same level as the hazard at one month in the model with no frailty and 73% below in the model with frailty correction.

## 5.3   Hysteresis in the log-Normal Hazard Functions

We shall now proceed to a detailed analysis of the log-normal hazard functions by splitting the whole dataset into a few subsamples. First, we split our sample according to the quartile of the covariates price decline and interest rate to which the spells (subjects) belong. Each covariate is split individually according to their own quartiles and the size of the underlying subsamples is 90-91 spells each. Next, we use the recession dummy as a factor variable in order to obtain separate fits of the hazard function and graphs for spells occurring or not during periods of economic recession. Note that for the whole dataset there are 114 spells of price decline occurring during economic recessions and 248 occurring otherwise. Wherever possible we show the hazard function fits for the log-normal with and without frailty correction.[14]

In Figure 7 we present the hazard curves obtained by quartile of the price decline covariate. For spells in the lower quartile of monthly stock market price decline there is no real distinction between the hazard curves during recessions and no recessions. The peak of both curves occurs at 2.8 months and their mean difference is only 0.003. After the peak there is a very slow decay of the hazard function which is not noticed by eye because the underlying time interval just spans 1 to 3 months. Anyway, the highest hazard rates occur precisely in this group which combines low percentages of monthly price decline with short lengths of spell duration. We may call this as frictional or transitory financial movements, i.e. periods wherein stock market prices adjust in order to synchronize with the whole financial system. Over this kind of spells, the probability of ending a stretch of financial market price decline increases as the spell length progresses and the stretch tends to be relatively short. This happens for spells lasting less than 3 months and a monthly price decline lower than 1.13%.

---

[14] For the price decline covariate, it was not possible to fit the frailty model for the 1st quartile. Likewise, for the covariate denoting the interest rate it was not possible to fit the frailty model both for the 1st and 3rd quartiles.

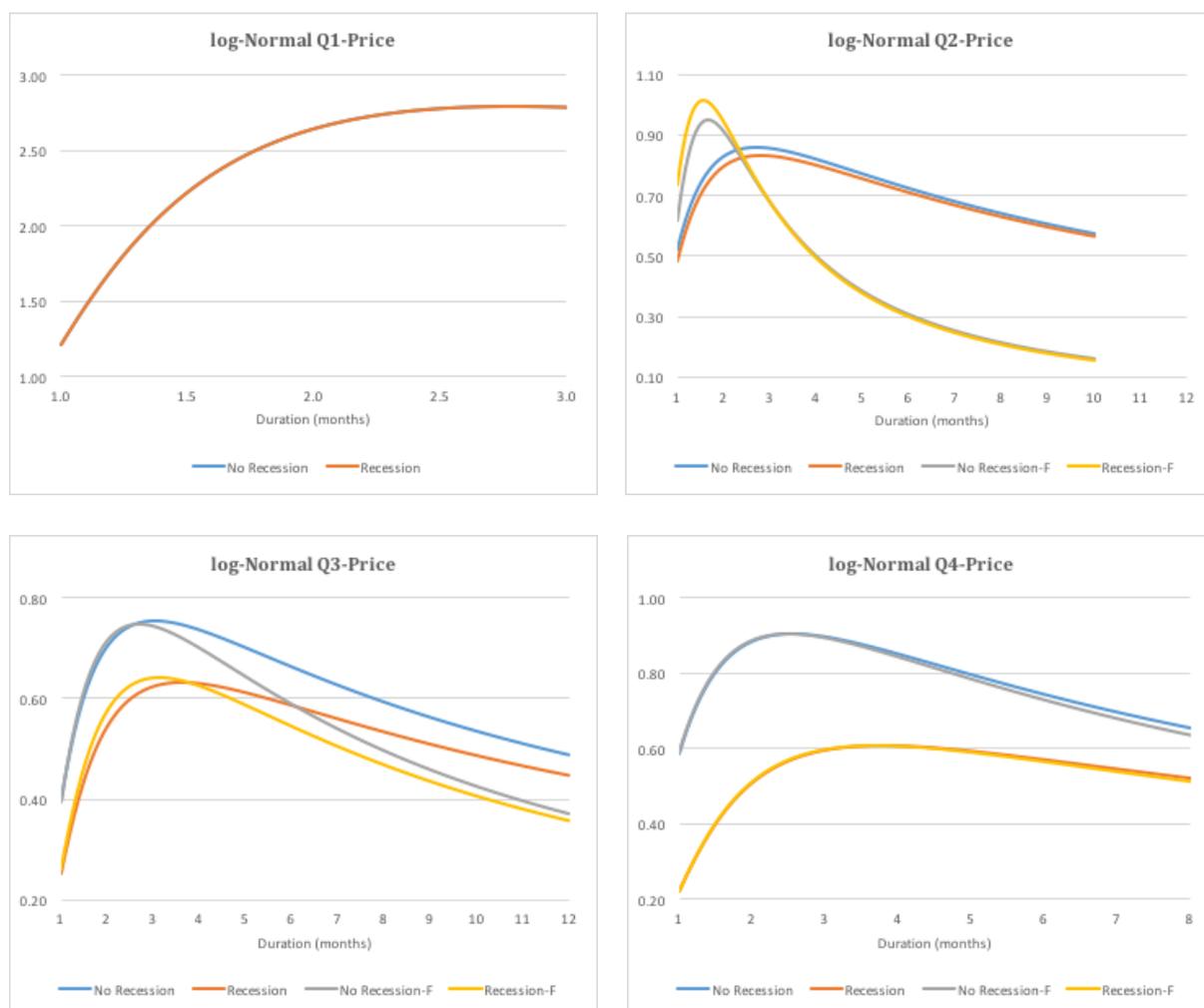

Figure 7. Hazard functions of the S&P500 duration time by quartile range of *Price Decline*, 1871-2016 (362 spells). Label F denotes frailty estimates

Turning now to the interquartile range of monthly stock market price declines we can see that during recessions the hazard curve tends to move away from the hazard curve during no recessions, especially in the left tail of the distribution. However, as time progresses the difference tends to vanish. In addition, when we move from the second to the third quartile range, the distinction between hazards during recessions or no recessions gets more pronounced in the left tail. The maximum holds at 2.7-2.8 months in the 2nd range and a little bit later, 3.0-3.6 months, in the 3rd range but at hazard rates substantially lower. A similar picture is obtained for the models estimated with frailty correction, with lower peak hazard rates occurring later during recession periods. For the interquartile range, the probability of ending a stretch of financial market price decline rises up to its maximum occurring at a length of 2.7-3.6 months, and falls after that point. This is the case of spells of any length and a monthly price decline between 1.13% and 3.49%. Such situation can occur, for example, when a mild frictional financial market breakdown turns to be a structural crisis after some time (3 months in our case) because the appropriate "treatment" was not applied or at least it was not applied efficiently. So, after the initial period of contagion, these spells exhibit signs of hysteresis or negative duration dependence, wherein as time progresses the probability of ending a spell of price decline falls.

Finally, for spells in the upper quartile of monthly stock market price decline we find the clearest evidence where hazard rates differ dramatically during recessions and no recessions. The peak during periods of recession holds at 3.8 months with a hazard rate of 0.60; during periods of no recession the peak occurs at 2.5 months with a hazard rate of 0.90, both in the model with and without frailty correction. Although spells in this range do not last beyond 8 months, the severity of the price decline can be tough. In this range hysteresis is clearly more pronounced during recessions than no recessions and the average monthly price decline is located between a minimum of 3.5% and a maximum of 18.2%.

To summarize, during recessive periods the turning point occurs between 2.8 and 3.8 months and the hazard rate falls monotonically from 2.79 to 0.61, as the decline of prices intensifies. During non-recessive periods the turning point occurs earlier, between 2.5 and 3 months, and the hazard rate falls non monotonically from 2.79 to 0.76. The results of the models estimated with frailty correction show a relatively similar pattern. During recessive periods the turning point holds at 1.5 to 3.7 months and the hazard rate falls monotonically from 1.01 to 0.61. On the other hand, during non-recessive periods the turning point occurs between 1.6 and 3 months and the hazard rate lies in the interval between 0.95 and 0.76, that is, the time interval is wider and the hazard range is shorter than in the model without frailty.

By turning point, we mean the moment in time where the hazard function reaches its peak and duration dependence changes from positive (increasing) to negative (decreasing). Positive duration dependence is associated with frictional spells of price decline and negative duration dependence is linked with hysteresis or increasing persistence. In the latter case, the longer the spell the lower the probability that it ends.

Figure 8 depicts the hazard curves obtained by quartile of the long-term interest rate covariate. The results are in some way opposite to those obtained for the price decline variable, in the sense that the similarity of the hazard curves for recessive and non-recessive periods now occurs in the upper quartile range. The profiles of the hazard functions remain pretty unchanged in the other quartile ranges, with higher failure rates occurring for non-recessive periods. Peaks occur more or less at the same time intervals as for the price decline figures. Overall, there is no outlandish pattern arising from the breakdown of the interest rate variable.

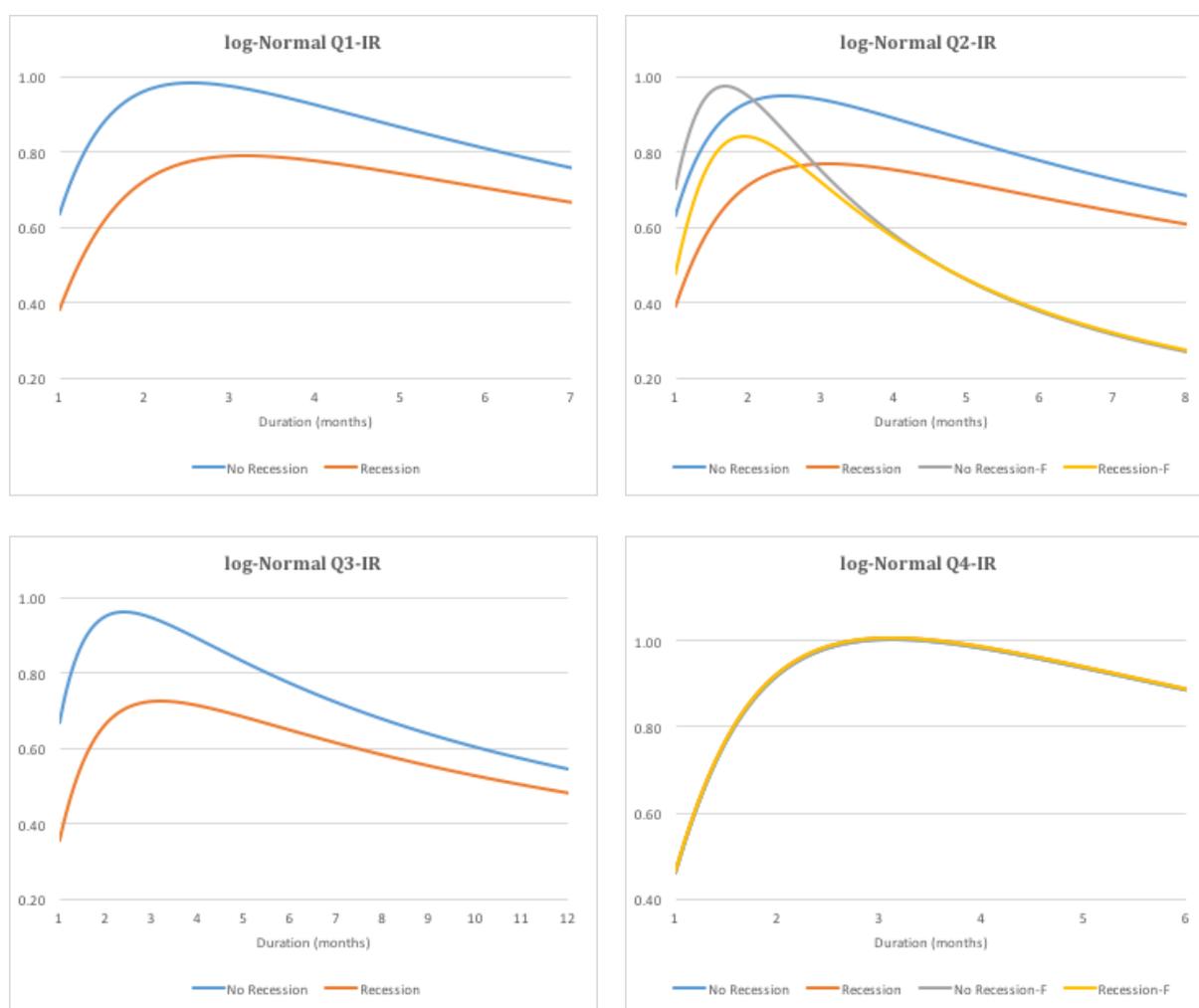

Figure 8. Hazard functions of the S&P500 duration time by quartile range of *Interest Rate*, 1871-2016 (362 spells). Label F denotes frailty estimates

The conclusion is that the probability of ending a spell duration of financial market dwindling increases up to about 2.5-4 months and decreases after that time. Shorter spells are more likely to end in a short time interval than longer spells. The latter are more frequent during economic recessions. The higher the price decline the longer is the spell duration and similarly with respect to long-term interest rates. Thus, monitoring of the initial characteristics of the financial market dwindling may provide some guidance about how long is expected to last this decline. It is, therefore, an important piece of information for economic and financial policy.

### 5.4   log-Normal Residuals

To complete the empirical section of the study, we present in this subsection a brief analysis of the residuals obtained by estimating the log-Normal regression with and without frailty correction. In survival analysis three types of residuals are in general computed: Cox-Snell, martingale-like and deviance residuals. Cox-Snell and martingale-like residuals are sometimes used for assessing the proportional hazards assumption and evaluate nonlinearities. However, they have a skewed and leptokurtic distribution which motivates the use of deviance residuals in order to achieve a more symmetric distribution for testing the accuracy of the model fit.

In Table 4 we present the main descriptive statistics of these residuals along with $\chi^2$ tests for autocorrelation (Breusch-Godfrey or BG) and heteroskedasticity (Breusch-Pagan-Godfrey or BPG). Neither the null of no autocorrelation nor the null of homoskedasticity in the residuals are rejected at standard levels in any of the tests. However, the null of Gaussian residuals is rejected in all cases (see the Jarque-Bera statistic) and the Cox-Snell and martingale-like residuals are characterized by skewness and excess kurtosis, as expected. Positive skewness occurs in the case of the Cox-Snell residuals and negative skewness occurs in the case of the martingale-like residuals. On the other side, deviance residuals are characterized by smaller values of skewness and kurtosis, also as expected. In this latter case, skewness remains slightly negative and the residual′s distribution tends to be mesokurtic. We may then conclude that robust estimation and, to some extent, frailty correction are in fact appropriate in our modeling process and the presence of outliers in the residuals is due to abnormal duration spells not entirely in line with the general pattern of the data.[15]

---

[15] The only outlier which was identified corresponds to the twelve-month spell that occurred in 1876/7, during the Long Depression.

| | Cox-Snell residuals | | | | Martingale-like residuals | | | | Deviance residuals | | | |
|---|---|---|---|---|---|---|---|---|---|---|---|---|
| | No frailty | | Frailty | | No frailty | | Frailty | | No frailty | | Frailty | |
| | stat | *p*-val. | stat | *p*-val. | stat | *p*-val. | stat | *p*-val. | stat | *p*-val. | stat | *p*-val. |
| Mean | 1.0194 | | 1.4692 | | -0.4692 | | -0.0194 | | 0.0887 | | 0.3423 | |
| Median | 0.7309 | | 1.0245 | | -0.0245 | | 0.2691 | | -0.0243 | | 0.2979 | |
| Maximum | 7.1789 | | 11.817 | | 0.9916 | | 0.9892 | | 2.7534 | | 2.6596 | |
| Minimum | 0.0108 | | 0.0084 | | -10.817 | | -6.1789 | | -4.0859 | | -2.9009 | |
| Std. Deviation | 1.1053 | | 1.7671 | | 1.7671 | | 1.1053 | | 1.2922 | | 1.0227 | |
| Skewness | 1.9733 | | 2.1554 | | -2.1554 | | -1.9733 | | -0.5532 | | -0.5159 | |
| Kurtosis | 8.0932 | | 9.2143 | | 9.2143 | | 8.0932 | | 2.4787 | | 2.4732 | |
| Jarque-Bera | 626.20 | 0.000 | 862.78 | 0.000 | 862.78 | 0.000 | 626.20 | 0.000 | 22.57 | 0.000 | 20.24 | 0.000 |
| Sum | 369.03 | | 531.86 | | -169.86 | | -7.03 | | 32.11 | | 123.91 | |
| Sum Sq. Deviation | 441.1 | | 1127.2 | | 1127.2 | | 441.1 | | 602.8 | | 377.6 | |
| *N* | 362 | | 362 | | 362 | | 362 | | 362 | | 362 | |
| *Breusch-Godfrey Serial Correlation LM Test* | | | | | | | | | | | | |
| $\chi^2(2)$ | 1.4921 | 0.4742 | 1.5718 | 0.4557 | 1.5718 | 0.4557 | 1.4921 | 0.4742 | 2.4612 | 0.2921 | 2.4323 | 0.2964 |
| *Heteroskedasticity Test: Breusch-Pagan-Godfrey* | | | | | | | | | | | | |
| $\chi^2(1)$ | 0.0571 | 0.8111 | 0.0485 | 0.8256 | 0.0485 | 0.8256 | 0.0571 | 0.8111 | 0.0246 | 0.8753 | 0.0061 | 0.9377 |
| Scaled expl. SS | 0.2008 | 0.6541 | 0.1975 | 0.6567 | 0.1975 | 0.6567 | 0.2008 | 0.6541 | 0.0183 | 0.8925 | 0.0045 | 0.9465 |

Table 4. Descriptive statistics (Cox-Snell, martingale-like and deviance residuals), serial correlation (BG Lagrange multiplier) and heteroskedasticity (BPG) tests for the residuals of the log-Normal regression with and without frailty, 1871-2016 ($N$ = 362 spells). $H_{0(BG)}$ = no serial correlation; $H_{0(BPG)}$ = no heteroskedasticity

Figure 9 shows the residuals for the log-Normal regression with and without frailty where, as we can see, a similar pattern is observed in all cases.

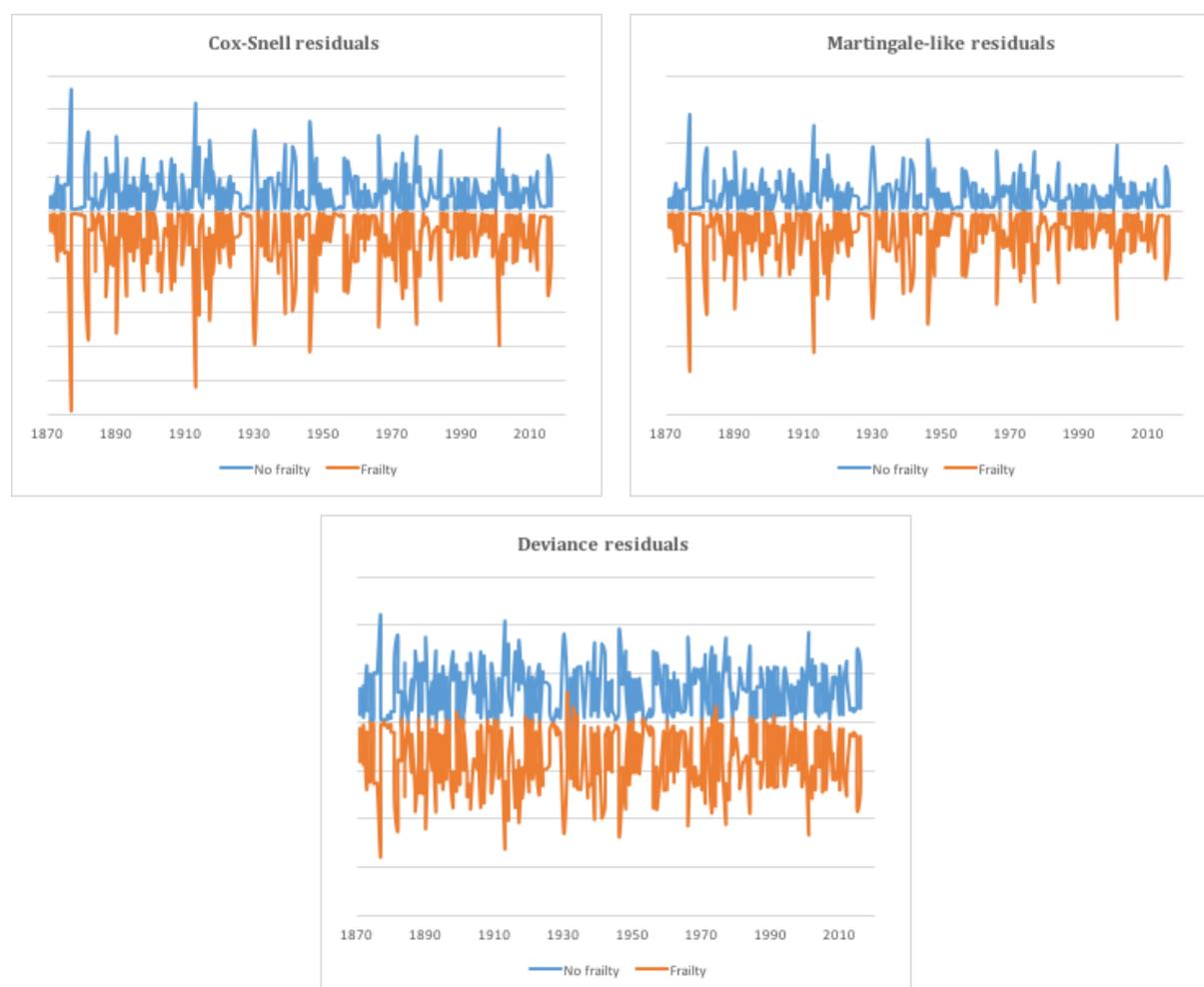

Figure 9. Residuals (Cox-Snell, martingale-like and deviance) of the log-Normal regression with and without frailty, 1871-2016 ($N$ = 362 spells). No frailty in blue and frailty in red. Scale was adjusted to facilitate the comparison between no frailty/frailty residuals and should be only interpreted in terms of magnitude, not in terms of residual′s sign

Finally, Figure 10 presents the Quantile-Quantile plots of the residuals against the Normal distribution. Cox-Snell and martingale residuals show significant deviations from the Gaussian in both tails denoting that PH, in our case, is actually not appropriate to explain tail events. However, deviance residuals only depart from the Gaussian for extreme positive values in the upper quartile of the distribution. As noted by Cleves *et al.* (2004), this type of deviations may be expected, and, thus, we may think that the log-Normal model adequately fits the data.

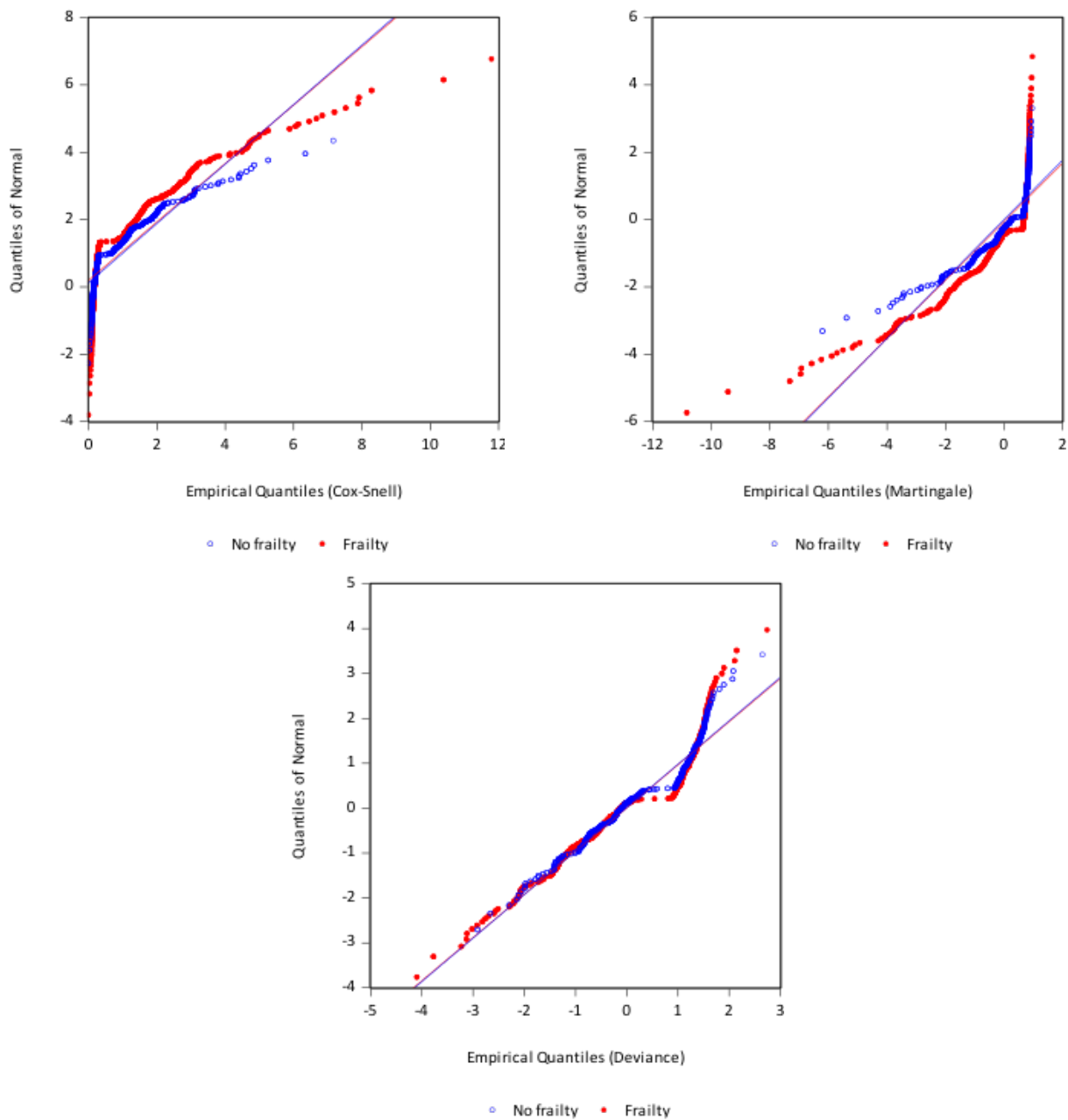

Figure 10. Quantile-Quantile plot for the residuals (Cox-Snell, martingale-like and deviance) of the log-Normal regression with and without frailty, 1871-2016 ($N = 362$ spells). No frailty in blue and frailty in red. Theoretical distribution is the Normal

Our results are in line with the notion that hysteresis, in the sense of negative duration dependence, is present in stock market price declines but this effect only arises after an initial period of market adjustment characterized by positive duration dependence, an issue that seems fundamental for designing economic and financial policies during cyclical crises.

# 6   Conclusions

Many studies in financial econometrics deal with the issue of volatility persistence using time series data but, to the best of our knowledge, no attempts have been made to model the duration of events that rule the behavior of the persistence of volatility itself. In stock markets, these events are related to price index declines and the existence of a large dataset without too much noisy data facilitates the task of building duration data. In our case, we use a monthly time series for US data spanning 145 years in order to obtain the duration of spells of stock market price decline, the intensity of the decline itself measured by the average percentage decrease per month, the average long-term nominal interest rate and a record of economic recession/no recession. Although other variables were also available, they did not prove to be statistically significant in our models.

Obviously, the nature of the spell depends on what we define as being the event of interest. In our study, the event of interest is the observation of an episode of no price decline in the stock market where each episode lasts a month. So, in our case a spell is the time window measured in months from the beginning to the end of a homogeneous period of consecutive months of price decline. In the words of Granger (2005), a spell would be a consecutive number of moments or episodes wherein stock prices put investors at risk of experiencing losses.

The literature on spell duration usually relies upon survival models and we also use this framework in our study. Models can be estimated under the assumption of proportional hazards or accelerated failure time. Although our results cannot discharge the assumption of proportional hazards, the main statistics related to the goodness-of-fit and model choice clearly endorse the use of accelerated failure time metrics. This is based on the AIC and SSR minimization. The models were estimated using robust standard errors in order to control heteroscedasticity. In order to account for omitted heterogeneity, we have also estimated our models with and without frailty correction. The best choice among the various specifications attempted falls on the log-normal parametric model. In fact, this model reaches the lowest AIC and SSR when estimated without frailty but there is no clear evidence that the frailty correction is actually necessary.

The results obtained are quite clear. During recessions, price decline spells tend to be 16-23% longer than during expansions. Spell's duration increase around 4% for every percentage point drop in the market price index and around 3% for every percentage point rise in interest rates.

The scale parameter falls between 0.45 and 0.56 indicating that the hazard function increases up to a peak and then decreases asymptotically to zero. The peak in the log-normal is reached 1.8-2.8 months after the outset of the spell with a hazard of 0.89 or higher. However, the overall figures conceal a variety of situations for different levels of the covariates where the peaks can be reached at about 4 months after the outset and the hazard falls with the quartile level of the covariate.

Naturally, other events could be used in order to define the spell of reference but for this study we wanted to keep the definition as easy to understand and as simple to compute as possible. Alternatives could use a different time span like, e.g., the time duration until losses are completely regained, which includes months of positive returns as well but this requires different definitions of what is the relevant event. For example, in this case the event occurs when the real price index recovers completely from its previous peak. Another alternative could include a semi-Markovian framework where the behavior going back one or more spells prior to the current one may influence the ongoing duration. Competing risks and multi-state models like, e.g. the inclusion of growth periods, their intensity, the occurrence of bubbles, etc. could also be attempted. Segmentation according to excess volatility is also a possible framework as well as economic crises associated with financial losses. Future work may also include the use of quantile-duration models.

Finally, two questions not (yet) answered in this study. First, can we really predict in advance that a market price decline is going to be short or long, i.e. just a market correction or a bear market? How much in advance could we predict this? Second, what kind of measures should stakeholders and decision-makers take in order to prevent that a market correction turns into a bear market?